\documentclass[twocolumn, twocolappendix]{aastex63}

\usepackage{comment}

\usepackage[italicdiff]{physics} 
\usepackage{amsmath}

\usepackage{graphicx}
\usepackage{float}

\usepackage{bigstrut}

\usepackage{makecell}

\newcommand{\m}[1]{\underline{\underline{#1}}}

\usepackage[encapsulated]{CJK}
\usepackage{ucs}
\usepackage[utf8x]{inputenc}
\usepackage{url}

\newcommand{\cntext}[1]{\begin{CJK}{UTF8}{bsmi}#1\end{CJK}}

\usepackage{enumitem}

\begin{document}


\title{Survival is not Enough: Dust Sputtering, Growth, and H$_2$ Formation in Galactic Winds}

\author[0000-0002-9235-3529]{Chia-Yu Hu (\cntext{胡家瑜})}
\affiliation{Institute of Astrophysics \& Department of Physics, National Taiwan University,\\ No. 1, Sec. 4, Roosevelt Rd., Taipei 10617, Taiwan}

\author[0000-0002-9235-3529]{Max Gronke}
\affiliation{Centre for Astronomy of Heidelberg University, Astronomisches Rechen-Institut, \\
M\"{o}nchhofstr. 12-14, 69120 Heidelberg, Germany}


\begin{abstract}

A substantial amount of dust is found in galactic halos extending far beyond the disks, the origin of which remains an open question.
Closely linked and equally puzzling is the detection of molecular gas in high-velocity galactic winds.
To address this, we present the first cloud-crushing simulations that self-consistently include non-equilibrium cooling and chemistry with dust growth and sputtering.
We find that surviving clouds naturally develop a two-phase structure, with a cold ($\sim 30$ K), dense core embedded in a warm ($\sim 10^4$ K), diffuse envelope.
However, the presence of a cold phase does not always lead to molecular winds.
While dust initially in the cloud largely survives in $10^6$~K winds, it is severely depleted by sputtering in hotter winds ($\gtrsim 10^7$~K).
Importantly, without dust growth, the dust-to-gas ratio (DGR) of the cloud declines rapidly, suppressing the formation of molecular hydrogen (H$_2$) and keeping the entrained cloud atomic, even in cases where the majority of the initial dust survives.
Nonthermal sputtering plays a subdominant role in all cases.
The entrained clouds develop high molecular fractions only when dust growth is enabled, provided the cloud densities are sufficiently high ($\gtrsim$ 10 -- 30 times the critical density for cloud survival).
Our results suggest that \textit{in situ} dust growth is essential to explain both the observed abundance of halo dust and the molecular gas in galactic winds.

\end{abstract}

\section{Introduction}\label{sec:intro}

Molecular clouds are the coldest and densest phase of the interstellar medium (ISM), 
with characteristic temperatures of 10 -- 50 K and densities of 100 cm$^{-3}$ \citep{Draine2011}.
They serve as the sites for star formation as seen not only in the Milky Way \citep{Lada2003} but also in galaxies in the local \citep{Kennicutt2012} and high-redshift Universe \citep{Tacconi2020}.
The main constituent of molecular clouds is molecular hydrogen (H$_2$), and the transition from atomic hydrogen gas (HI) to molecular gas has therefore been extensively studied \citep{Krumholz2008, Sternberg2014}.
This transition depends critically on the presence of interstellar dust, whose importance is twofold.
First, dust grains act as catalysts for H$_2$ formation, providing the surfaces where hydrogen atoms can meet and bond much more efficiently than the gas-phase route. Secondly, dust provides important shielding against the far-ultraviolet (FUV) radiation that would otherwise efficiently photodissociate H$_2$. 

There is growing observational evidence that both dust and molecular gas exist well beyond galactic disks. 
Reddening measurements of background quasars against foreground galaxies have found a substantial amount of dust in the circumgalactic medium (CGM) of galaxies and even in the intergalactic medium (IGM) on megaparsec (Mpc) scales \citep{Menard2010, Peek2015}.
Molecular gas and polycyclic aromatic hydrocarbon (PAH) have also been observed around the Brightest Central Galaxies (BCGs) of galaxy clusters as filamentary structures in the vicinity of hot, X-ray-emitting intracluster gas \citep{Salome2006, Donahue2011, Russell2019, Olivares2019}.
The origin of this dusty, molecular gas is still an open question: is it formed in situ beyond the galaxies, or is it transported from the galaxies?

Observations of galactic outflows provide compelling evidence that outflows (or winds) are a major carrier of both dust and molecular gas (see \citealp{Veilleux2020} for an extensive review).
Thermal emission from dust has been directly detected in galactic winds from star-forming galaxies \citep{McCormick2013, McCormick2018}. 
Molecular outflows are frequently observed in a variety of galactic environments, from nearby starburst galaxies driven primarily by supernovae (SNe) \citep{Bolatto2013a, Leroy2015, Zschaechner2018, Krieger2021, Li2025}
to Active Galactic Nuclei (AGN)-active galaxies such as ultra-luminous infrared galaxies (ULIRGs), quasars, or Seyfert galaxies \citep{Feruglio2010, Sturm2011, Combes2013, GonzalezAlfonso2017, Lutz2020}.
Molecular outflows are also observed in the high-redshift Universe, both in dusty, star-forming galaxies \citep{Spilker2020, Spilker2020a} and in galaxies with active AGN activities \citep{HerreraCamus2020, Spilker2025}.

However, the presence of the cold, dusty molecular gas expelled/uplifted by the hot, energetic winds poses a puzzling problem.
Theoretical models suggest that dust grains entrained in hot environments such as galactic winds should be rapidly destroyed by thermal sputtering \citep{Ferrara2016}.
If dust is indeed destroyed, the existence of the molecular phase becomes equally puzzling, as the lack of dust removes both the formation catalyst and the shielding agent for molecular gas. 
Resolving this tension between the observed ubiquity of cold outflows and the theoretical predictions of rapid destruction is crucial for understanding the gas cycle in and around galaxies.

Recently, high-resolution simulations of a hot galactic wind impinging on a colder cloud (often referred to as ``cloud-crushing'' simulations) have provided new theoretical insights into how cold gas can persist in these hostile environments. 
The key realization, first pointed out by \citet{Gronke2018}, was that ``cool'' clouds ($T\sim 10^4$~K) can survive and even grow while being entrained in hot galactic winds, provided that radiative cooling is efficient enough to condense the mixing-layer gas and acquire the momentum. 
Extensive work has subsequently established this result in various regimes (see \citealp{FaucherGiguere2023, GronkeSchneider2026} for comprehensive reviews).
However, many open questions remain, particularly regarding the phase transition required for H$_2$ formation. 
Molecular clouds reside at temperatures of a few tens of Kelvin, significantly colder than the $10^4$~K temperature floor typically adopted in most cloud-crushing simulations. 
Reaching these low temperatures is critical because the formation efficiency of H$_2$ depends on the sticking coefficient of hydrogen atoms onto the surfaces of dust grains, which drops sharply at higher temperatures \citep{Hollenbach1979}.

Recent studies have begun to probe these regimes, though each with specific limitations. \citet{Farber2022} investigated the entrainment of $10^3$~K clouds, including dust sputtering; they found that large clouds can survive the cloud-wind interaction by effectively avoiding exposure to the hot gas. 
However, their cooling function did not extend to the temperatures typical of molecular gas. 
On the other hand, \citet{Girichidis2021} adopted non-equilibrium cooling and chemistry to explicitly study H$_2$ formation. 
However, the simulations were not evolved long enough to reach the fully entrained phase of the cloud, where continued growth may substantially alter the molecular content. In addition, their model did not include dust sputtering, which could overestimate H$_2$ formation.
Other works have focused on the evolution of dust. 
\citet{Chen2024} included both dust growth and sputtering and found that cold gas at molecular temperatures can form with significant dust surviving as it is protected by a warm envelope.
\citet{Richie2024} modeled dust sputtering with a temperature floor of $10^4$ K and found that dust can survive even in the absence of a distinct cold phase. 
While all these simulations provide crucial insights, they represent pieces of a larger puzzle. 
To date, no cloud-crushing simulations have integrated dust evolution (with growth and sputtering) with a time-dependent H$_2$ chemistry simultaneously, all of which are crucial components to tackling the problem.
This is the gap our study aims to fill.

In this work, we conduct the first cloud-crushing simulations that self-consistently follow the complex interplay between radiative cooling, cloud-wind interaction, dust sputtering and growth, and H$_2$ formation.
This allows us to obtain a more complete picture of how dust and molecular clouds are entrained by a hot, fast-moving wind.
We shall demonstrate that while dust that originally resides in the cloud can survive in $10^6$~K winds, it is rapidly depleted by sputtering in the $10^7$~K winds.
More importantly, regardless of the degree of dust survival, the dust-to-gas ratio (DGR) of the cloud is reduced by more than an order of magnitude when dust growth is disabled, hindering H$_2$ formation and leading to HI-dominated clouds.
Entrained clouds with high H$_2$ fractions are only possible with dust growth, provided that the cloud densities are sufficiently high.

This paper is organized as follows. 
In Section~\ref{sec:methods}, we describe the numerical methods and our simulation setup. In Section~\ref{sec:results}, we present the results which we discuss in Section~\ref{sec:discussion} before concluding in Section~\ref{sec:conclusion}.

\section{Numerical Methods} \label{sec:methods}

\subsection{Simulation Code}

We use the public version of {\sc Gizmo} \citep{Hopkins2014}, a code for gravity and hydrodynamics built on the {\sc Gadget-2} code \citep{Springel2005}, though we do not include self-gravity in this work.
We adopt the meshless finite-mass (MFM) solver for hydrodynamics \citep{Hopkins2015},
a variation of the meshless Godunov method \citep{Gaburov2011}.
In addition, we include non-equilibrium cooling and chemistry developed by \citet{Glover2007} and \citet{Glover2012a} that are widely adopted in ISM-scale simulations.
The chemistry network includes H$_2$ formation via the dust channel, H$_2$ destruction via photodissociation, collisional dissociation, and cosmic ray ionization,
and recombination in the gas phase and on dust grains.
The gas-phase H$_2$ formation is only relevant at extremely low DGRs \citep{Glover2003} and thus can be safely neglected in our case.
Cooling and heating are calculated directly from individual non-equilibrium chemical abundances.
Shielding against the far-ultraviolet (FUV) radiation is treated with an empirical relation presented in Appendix \ref{app:NHnH},
which we find more accurate than the conventional Sobolev approximation.
We assume that the cloud is far enough from the host galaxy and thus is unaffected by the starlight from the latter. 
As such,
the unattenuated FUV radiation field strength comes from the cosmic UV background,
which we set to be $G_0 = 3.2\times 10^{-3}$ in Habing units \citep{Habing1968} following \citet{Hu2017}.
However, we also run a set of simulations with $G_0 = 1.7$ (the Draine field), appropriate for clouds that are still very close to the disk.
Photoionization heating from the cosmic UV background is not included.
As \citet{Chen2024} pointed out, the condition for cloud survival coincides with the column density where the cloud becomes self-shielded against the cosmic ionizing radiation.
As such, we assume that all clouds are sufficiently well-shielded.
Although this approximation breaks down when clouds are dispersed by fluid instabilities, these disruptive cases are not the focus of this work.

\begin{deluxetable*}{ccccccccc}
	\tablecaption{
		\label{tab:params}
        Parameters in our simulation setup.
	}
	\tablewidth{0pt}
	\tablehead{
		\colhead{$T_{\rm w}$}   &
		\colhead{$n_{\rm c}$}  &
		\colhead{$n_{\rm w}$}   &
		\colhead{$v_{\rm w}$}   &
		\colhead{$P$}   &
		\colhead{$t_{\rm cc}$}   &
		\colhead{$n_{\rm c}^\prime = n_{\rm c} / n_{\rm crit}$ }  &
		\colhead{box size}  &
		\colhead{fate of cloud}  \\
		\colhead{[K]} &
		\colhead{$[{\rm cm^{-3}}]$} &
		\colhead{$[{\rm cm^{-3}}]$} &
		\colhead{$[{\rm km~s^{-1}}]$} &
		\colhead{$[{\rm cm^{-3}~K}]$} &
		\colhead{$[{\rm Myr}]$} &
		\colhead{} &
		\colhead{[kpc$^3$]} &
		\colhead{} 
	}
	\startdata
	$10^6$   & $0.01$  & $        10^{-4}$ &   200	&   $        10^{2}$  & $5.0$ & $0.333$ & $2^2\times 14$ &	destroyed \\
	$10^6$   & $ 0.1$  & $        10^{-3}$ &   200	&   $        10^{3}$  & $5.0$ & $3.33 $ & $2^2\times 14$ &	survived  \\
	$10^6$   & $ 0.3$  & $3\times 10^{-3}$ &   200	&   $3\times 10^{3}$  & $5.0$ & $10.0 $ & $2^2\times 14$ &	survived  \\
	$10^6$   & $   1$  & $        10^{-2}$ &   200	&   $        10^{4}$  & $5.0$ & $33.3 $ & $2^2\times 14$ &	survived  \\
	\hline
	$10^7$   & $    1$ & $        10^{-3}$ &   620	&   $        10^{4}$  & $5.0$ & $0.351$ & $4^2\times 32$ &	destroyed \\
	$10^7$   & $   10$ & $        10^{-2}$ &   620	&   $        10^{5}$  & $5.0$ & $3.51 $ & $4^2\times 32$ &	survived  \\
	$10^7$   & $   30$ & $3\times 10^{-2}$ &   620	&   $3\times 10^{5}$  & $5.0$ & $10.0 $ & $4^2\times 32$ & survived  \\
	$10^7$   & $ 10^2$ & $        10^{-1}$ &   620	&   $        10^{6}$  & $5.0$ & $35.1 $ & $4^2\times 32$ & survived  \\
	\enddata
	\tablecomments{
    (1) $T_{\rm w}$ : wind temperature,
    (2) $n_{\rm c}$ : cloud density (hydrogen number density),
    (3) $n_{\rm w}$ : wind density,
    (4) $v_{\rm w}$ : wind velocity,
    (5) $P$         : thermal pressure,
    (6) $t_{\rm cc}$: cloud-crushing time,
    (7) $n_{\rm c}^\prime$: cloud density normalized by the critical density,
    (8) the dimensions of the simulation domain are identical along the $x$- and $y$-directions, while the $z$-direction is significantly extended. 
    The last column indicates whether the cloud survives or gets dispersed in the wind at the end of the simulation.
	All simulations adopt an initial cloud temperature $T_{\rm c} = 10^4~$K and a cloud radius $r_{\rm c} = 100~$pc, and the cloud is resolved with $10^5$ gas particles.
	}
    \vspace{-0.5cm}
\end{deluxetable*}

We adopt the dust evolution model from \citet{Hu2023}.
Dust is treated as a passive scalar with a mass of $m_{\rm d}$ 
associated with a gas particle with a mass of $m_{\rm g}$.
The local dust-to-gas ratio is therefore $Z_{\rm d} = m_{\rm d} / m_{\rm g}$.
While the gas particle mass remains constant, the dust mass evolves with time due to destruction via thermal and nonthermal sputtering following \citet{Hu2019a} and formation via dust growth:
\begin{equation}
    \frac{\mathrm{d}m_{\rm d}}{\mathrm{d}t} = (1 - f_{\rm d}) \frac{m_{\rm d}}{t_{\rm grow}} -\frac{m_{\rm d}}{t_{\rm sput}} 
\end{equation}
where 
$m_{\rm d}$ is the dust mass associated with the gas particle, 
$t_{\rm sput}$ is the sputtering time, $t_{\rm grow}$ is the growth time, and $f_{\rm d}$ is the local dust-to-metal ratio.
We assume a non-evolving grain size distribution following the MRN distribution \citep{Mathis1977} with the minimum and maximum sizes of $a_{\rm min} = 0.005 \mu m$ and $a_{\rm max} = 0.25 \mu m$, respectively. 
Therefore,
the grain-size integrated sputtering time can be expressed as
\begin{eqnarray}\label{eq:t_sput}
t_{\rm sput} &=& \frac{m_d}{\dot{m_d}} 
= \frac{a_{\rm eff}}{3 n Y_{\rm tot}} \nonumber\\
&\approx& 10 ~{\rm kyr}\ \Big(\frac{a_{\rm eff}}{0.035 \mu m}\Big) \Big(\frac{n}{{\rm cm}^{-3}}\Big)^{-1} 
\Big(\frac{10^{6} Y_{\rm tot}}{\rm \mu m \ yr^{-1}cm^3}\Big)^{-1},\nonumber\\ \end{eqnarray}
where $n$ is the hydrogen number density, $a_{\rm eff}$ is the effective grain size, and $Y_{\rm tot}$ is the erosion rate including thermal and nonthermal sputtering from \citet{Nozawa2006}.
The effective grain size is defined as 
$a_{\rm eff} \equiv \langle a^3 \rangle / \langle a^2 \rangle$
where the bracket represents the grain-size average 
$\langle Q \rangle = \int_{a_{\min}}^{a_{\max}} Q f(a) da$ for a given quantity $Q$
and $f(a) \propto a^{-3.5}$ is the MRN probability distribution function.
Our adopted range of grain size leads to $a_{\rm eff} = 0.035 \mu m$.
Thermal sputtering depends on gas temperature and density and is only effective in the hot gas if $T \gtrsim 10^6$~K,
while nonthermal sputtering depends on gas density and the dust-gas relative velocity $v_{\rm rel}$
and is effective in strong shocks when $v_{\rm rel} \gtrsim 100~\text{km/s}$.
Both thermal and nonthermal sputtering, when active, have characteristic erosion rates $\sim 10^{-6} {\rm \mu m \ yr^{-1}cm^3}$.
We calculate $v_{\rm rel}$ by integrating the equation of motion of dust
that accounts for direct collision and plasma drag,
ignoring the effect of betatron acceleration (as our system is not magnetized):
${dv_{\rm rel}} / {dt} = - v_{\rm rel} / t_{\rm drag}$
where 
\begin{eqnarray}\label{eq:t_drag}
t_{\rm drag} \approx 21~\text{kyr}  \Big( \frac{a_{\rm eff}}{0.035\mu m} \Big) \Big( \frac{n}{ \text{cm}^{-3} }\Big)^{-1} \Big( \frac{T}{ 10^6\text{K} }\Big)^{-0.5}.
\end{eqnarray}
The equation of motion is co-evolved with sputtering with the subcycling scheme described in \citet{Hu2019a}.
For dust growth,
the grain-size integrated growth timescale can be expressed as 
\begin{eqnarray}\label{eq:t_grow}
t_{\rm grow} 
&\approx& 1.5~\text{Gyr}~\Big(\frac{n}{ \text{cm}^{-3} }\Big)^{-1} \Big( \frac{T}{100\text{K}} \Big)^{-0.5}  \nonumber\\
&& \Big( \frac{a_{\rm eff}}{0.035 \mu\text{m}} \Big) (Z^{\prime} \alpha_s)^{-1}. 
\end{eqnarray}
where $Z^{\prime}$ is the normalized gas-phase metallicity ($Z^{\prime} = 1$ for solar metallicity)
and $\alpha_s$ is the sticking coefficient.
Dust growth is only efficient in cold gas, where the gas thermal velocity is low enough for metals to stick onto the grain surfaces, though the exact mechanism and temperature dependence are still uncertain.
For simplicity, we set $\alpha_s = 1$ for $T < 300$~K and $\alpha_s = 0$ otherwise, following \citet{Zhukovska2016}.

The evolving DGR is coupled to the chemistry in every timestep, affecting both H$_2$ formation and radiation shielding.
Note that the H$_2$ formation time on dust at $T\lesssim 100$~K is 
\begin{equation}\label{eq:t_H2}
	t_{\rm H_2} \approx 1.5~{\rm Gyr} ~Z^{\prime -1}_{\rm d} \Big(\frac{n}{{\rm cm}^{-3}}\Big)^{-1}  
\end{equation}
where $Z^\prime_{\rm d} = Z_{\rm d} / Z_{\rm d,MW}$ is the dust-to-gas ratio normalized to the Milky Way value ($Z_{\rm d,MW} = 5.4\times 10^{-3}$).
Coincidentally, 
$t_{\rm grow} \sim 	t_{\rm H_2} $ when $Z^{\prime} = Z^{\prime}_{\rm d} = 1$.

We include sub-grid turbulent diffusion of the chemical and dust abundances
using the metal diffusion module in {\sc Gizmo}\citep{Hopkins2018a,Colbrook2017}.
The diffusion coefficient is set to be one based on the \textit{a priori} turbulence simulations \citep{Hu2020}.
We note that it is necessary to include sub-grid turbulent diffusion in Lagrangian codes.
Without it, H$_2$ formation would be artificially limited to the original cloud particles, as most cooled wind particles are essentially dust-free due to efficient sputtering in the wind.

\subsection{Simulation Setup}

We conduct a suite of cloud-crushing simulations.
The initial conditions consist of a spherical cloud 
embedded in a hotter, lower-density background ``wind'' 
moving at a constant wind velocity $v_{\rm w}$.
The cloud and the wind are in pressure equilibrium.
All simulations have the same cloud radius $r_{\rm c} = 100~$pc\footnote{As stated below, this is without loss of generality since the evolution is self-similar and quantities can be rescaled accordingly \citep[see section 2.1.2 in][]{2025MNRAS.544.4621D}.} 
and cloud temperature $T_{\rm c} = 10^4~$K.
We run two sets of simulations with different wind temperatures, $T_{\rm w} = 10^6$~K and $10^7$~K, traveling at $v_w = 200~{\rm km~s^{-1}}$ and $620~{\rm km~s^{-1}}$, respectively,
corresponding to a Mach number $\mathcal{M} \sim 1.3$.
Previous cloud-crushing simulations have identified a crucial criterion for cloud survival,
which occurs when the timescale for radiative cooling in the cloud-wind interface is shorter than 
the cloud-crushing time 
\begin{eqnarray}\label{eq:t_cc}
t_{\rm cc} 
&=& \chi^{1/2} r_{\rm cl} / v_{\rm wind}\nonumber\\
&\approx& 
5~{\rm Myr} ~\left( \frac{\chi}{100} \right)^{1/2} \Big(\frac{r_{\rm cl}}{\rm 100~pc} \Big) \Big(\frac{\rm 200~km~s^{-1}}{v_{\rm wind}} \Big) 
\end{eqnarray}
where $\chi$ is the density contrast between the cloud and wind \citep{Klein1994}.
This leads to a critical radius for cloud survival for a given cloud density
\citep{Gronke2018,Gronke2020},
or,
equivalently,
a critical cloud density for cloud survival for a given $r_{\rm c}$
\begin{eqnarray}\label{eq:ncrit}
	n_{\rm crit} \approx 0.03 ~{\rm cm^{-3}} 
	\frac{T_{\rm c,4}^{1/2}}{ r_{\rm c,2} \Lambda_{\rm mix,-21.4} }	
	\left( \frac{\mathcal{M}}{1.5} \right)
	\left( \frac{\chi}{100} \right)^{1/2}
\end{eqnarray}
where 
$T_{\rm c,4} \equiv T_{\rm c} / (10^4{\rm K})$, 
$r_{\rm c,2} \equiv r_{\rm c} / (10^2{\rm pc})$, 
$ \Lambda_{\rm mix,-21.4}	\equiv \Lambda_{\rm mix} / (10^{-21.4}{\rm erg~cm^3~s^{-1}}) $, and
$\mathcal{M} = v_{\rm w} / c_{\rm s,w}$ is the Mach number in the wind.
Our Eq.~\ref{eq:ncrit} is directly converted from \citet{Chen2024} (see their Eq.~11).
We systematically vary the normalized cloud density 
$n^\prime_{\rm c} = n_{\rm c} / n_{\rm crit} = $ 0.3, 3, 10, and 30.
The wind density is scaled accordingly as $n_{\rm w} = n_{\rm c} / \chi$.
Cooling at $T> 0.6 T_{\rm w}$ is switched off mimicking a volumetric heating source and to avoid wind cooling.
The normalized metallicity and DGR are set to be $Z^\prime = Z^\prime_{\rm d} = 1$.
Hydrogen is initially fully ionized.
To track the cloud-wind mixing,
we initialize a passive scalar field $\phi = 1$ in the cloud and $\phi = 0$ in the wind.
The passive scalar is subject to sub-grid diffusion similar to the chemical and dust abundances.
The simulation parameters are summarized in Table~\ref{tab:params}.

\begin{figure}
	\centering
	\includegraphics[trim=2cm 1cm 2cm 1cm, clip, width=1\linewidth]{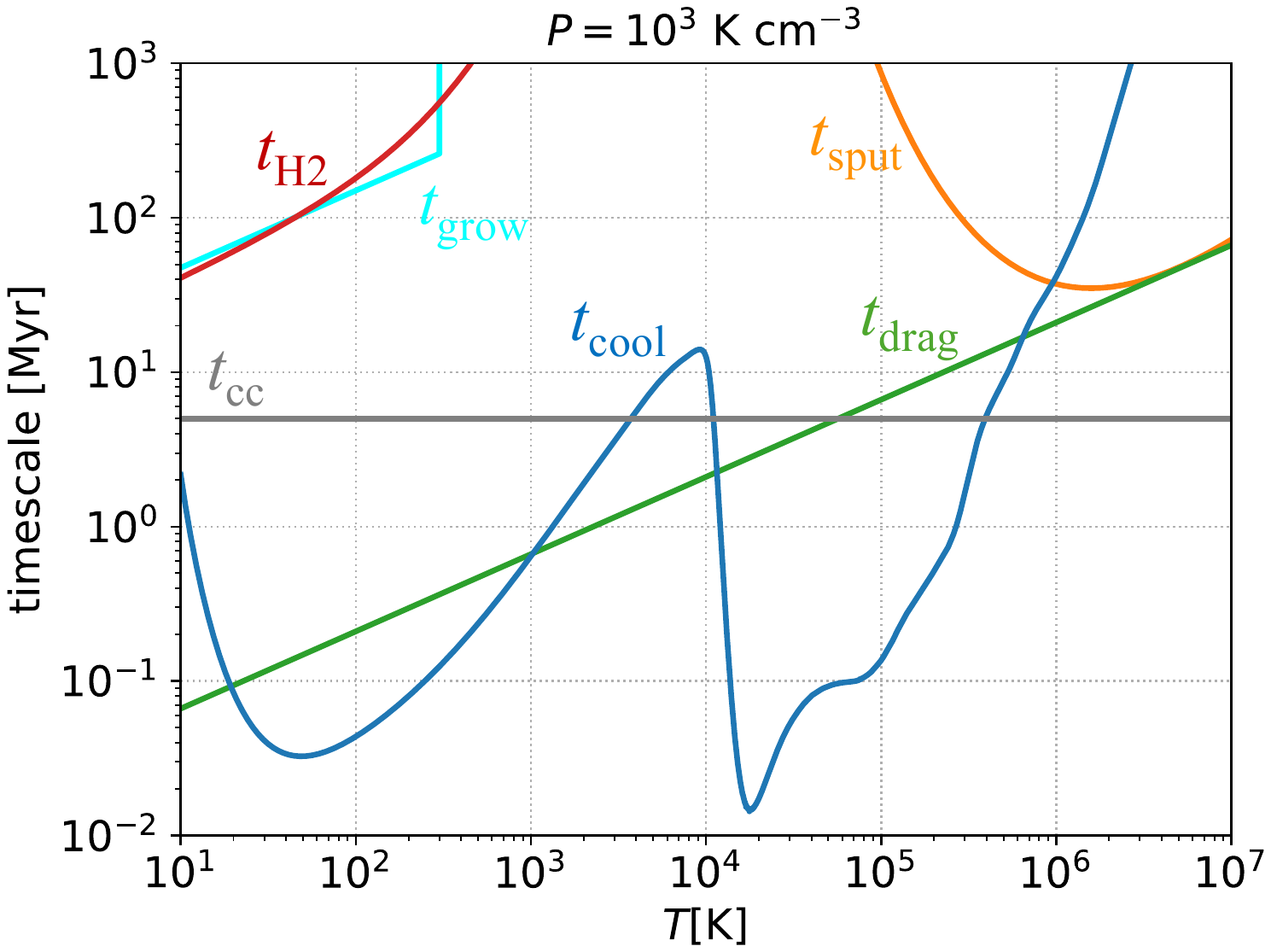}
	\caption{
    Various characteristic timescales as a function of gas temperature under a constant pressure $P = 10^3~{\rm K~cm^{-3}}$, including the 
    radiative cooling time in collisional ionization equilibrium ($t_{\rm cool}$),
    dust-gas drag time ($t_{\rm drag}$, Eq.~\ref{eq:t_drag}),
    dust growth time ($t_{\rm growth}$, Eq.~\ref{eq:t_grow}),
    H$_2$ formation time ($t_{\rm H_2}$, Eq.~\ref{eq:t_H2}),
    and cloud-crushing time ($t_{\rm cc}$, Eq.~\ref{eq:t_cc}).
	}
	\label{fig:timescale}
\end{figure}

Fig.~\ref{fig:timescale} shows the aforementioned timescales as functions of gas temperature, assuming that the different gas phases are in pressure equilibrium at $P = 10^3~{\rm K~cm^{-3}}$, meaning the gas density is inversely proportional to $T$.
Sputtering and drag occur primarily in the hot gas ($T \gtrsim 10^6$~K), while dust growth and H$_2$ formation take place in the cold gas ($T \lesssim 10^2$~K).
Radiative cooling occurs rapidly except at the peak near $T\sim 10^4$~K which is referred to as the `cooling bottleneck' in \citealp{Chen2024}).
Within their most active temperature ranges, these processes all occur within a few $t_{\rm cc}$ at this chosen pressure.
Except for the cloud-crushing time, which is density-independent, all other timescales scale inversely with density as they originate from collisional processes.
As such,
for clouds with pressures exceeding $P = 10^3~{\rm K~cm^{-3}}$ (as explored in our simulations), we expect all these physical processes to occur within a few $t_{\rm cc}$ or even significantly faster.

For each simulation setup, we run four different dust evolution models summarized in Table~\ref{tab:dustprocess}:
\begin{enumerate}[wide, labelwidth=!, labelindent=4pt]
\item our fiducial model with sputtering and dust growth,
\item a sputtering-only model with dust growth disabled,
\item a model with nonthermal sputtering switched off, 
\item a constant-DGR model with no dust evolution.
\end{enumerate}

The simulation domain is a box elongated along the wind direction.
The box size is (2 kpc)$^2 \times$ 14 kpc for the $T_{\rm w} = 10^6~$K wind
and  (4 kpc)$^2 \times$ 32 kpc for the $T_{\rm w} = 10^7~$K wind,
large enough to enclose the elongated cloud at late times.
The boundary conditions are periodic in all directions.
We do not adopt the cloud-tracking technique as the MFM method is Galilean-invariant by construction.
All our simulations are run for $24 ~t_{\rm cc}$.
In all simulations, the cloud is represented by $N_{\rm c} = 10^5$ equal-mass particles, randomly distributed within the cloud radius.
The wind is represented by particles distributed in a Cartesian grid.
We adopt the same particle mass for the wind and cloud.
Convergence tests are presented in Appendix~\ref{app:conv}.

The gas temperature in the simulations has a broad distribution spanning orders of magnitude.
We shall refer to 
gas with $T > 10^5$~K as \textit{wind}
and gas with $T < 10^5$~K as \textit{cloud} 
(which can fragment into several clouds).
The cloud is 
further divided into the 
\textit{warm} ($10^3 < T/{\rm K} < 10^5$) phase 
and the
\textit{cold} ($T < 10^3$~K) phase.

\begin{deluxetable}{lccc}
	\tablecaption{
		\label{tab:dustprocess}
	}
	\tablehead{
		\colhead{model}   &
		\colhead{\makecell[cl]{thermal\\sputtering}}  &
		\colhead{\makecell[cl]{nonthermal\\sputtering}}   &
		\colhead{\makecell[cl]{dust\\growth}}  
	}
	\startdata
	\makecell[cl]{sput. $+$ growth\\(fiducial)}  & o  & o  & o    \\
	growth off                  & o  & o  & x    \\
	nth. sput. off              & o  & x  & o    \\
	const. DGR                  & x  & x  & x    \\
	\enddata
	\tablecomments{
    Included processes in four different dust evolution models.
	}
    \vspace{-0.5cm}
\end{deluxetable}

\section{Results} \label{sec:results}

\subsection{Cloud Growth and Entrainment}

\begin{figure}
	\centering
	\includegraphics[trim=0cm 0cm -1cm 0cm, clip, width=1\linewidth]{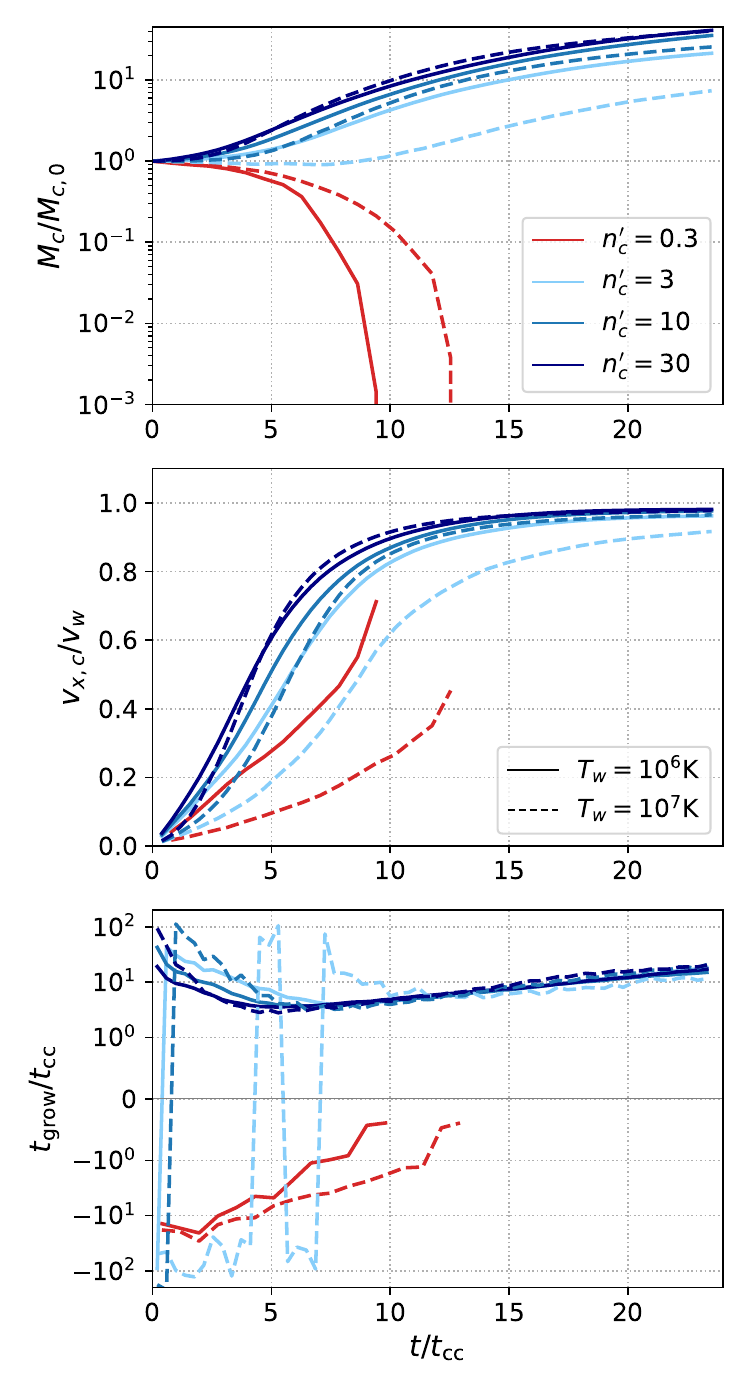}
	\caption{
        Global properties of the cloud (defined as gas with $T < 10^5$ K) as a function of time (in units of the cloud-crushing time, $t_{\rm cc}$)
        for different initial cloud densities ($n_{\rm c}$) and wind temperatures ($T_{\rm w}$).
		\textit{Top}: 
        Total mass of the cloud in units of the initial cloud mass ($M_c / M_{c,0}$).
        \textit{Middle}:
        Mass-weighted cloud velocity in the direction of the wind, normalized to the wind velocity ($v_{x, c} / v_{\rm w}$).
        \textit{Bottom}:
        Cloud growth timescale, defined as $t_{\rm grow} \equiv M_c / \dot{M}_c$ (where $\dot{M}_c$ is the mass growth rate of the cloud),
        in units of $t_{\rm cc}$.
		Clouds with an initial density higher than the critical density ($n_{\rm crit}$) grow and accelerate via cooling-driven accretion.
	}
	\label{fig:fig_warmcold_time_evol_single}
\end{figure}

\begin{figure*}
	\centering
	\includegraphics[width=1\linewidth]{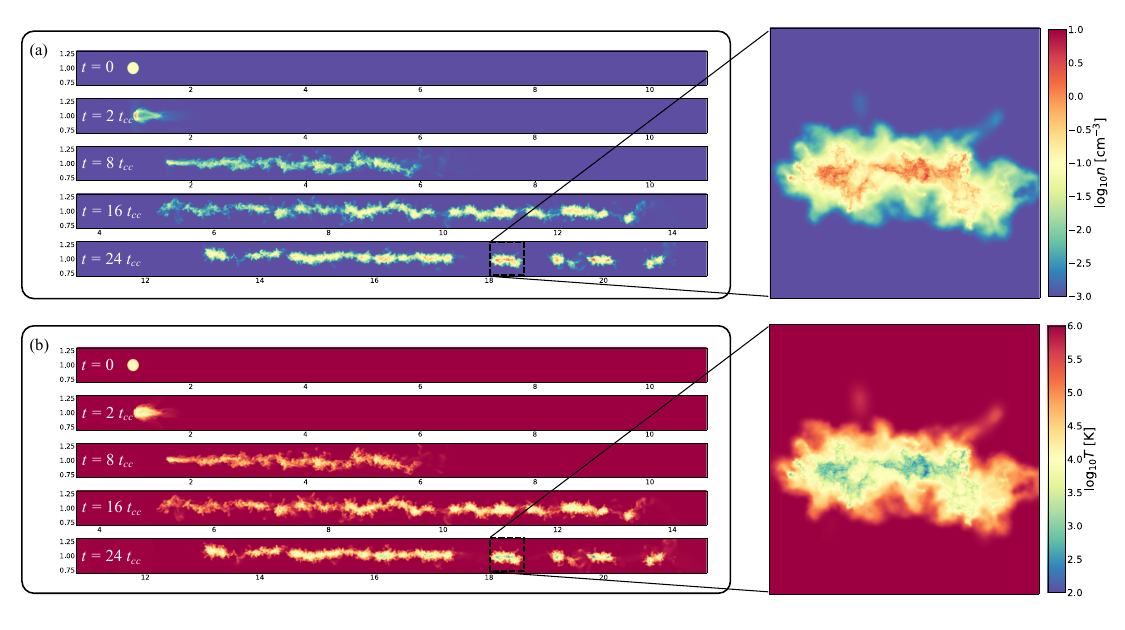}
	\caption{
		Visualization of the simulation with $T_{\rm w} = 10^6~{\rm K}$, $n_c = 3.3~n_{\rm crit}$ 
		at $t = $ 0, 2, 8, 16, and 24 $t_{\rm cc}$ from top to bottom.
		Panel (a) shows the hydrogen number density while panel (b) shows the gas temperature. 
		The right panels show zoom-in images of a cloud highlighted 
		in the dashed squares on the left panels at $t = 24~t_{\rm cc}$,
		demonstrating the two-phase ($T\sim 10^4$ K and $T \lesssim 10^2$ K) structure of the clouds.
	}
	\label{fig:maps_n_T}
\end{figure*}

Fig.~\ref{fig:fig_warmcold_time_evol_single} shows 
the global properties of the cloud as a function of time (in units of $t_{\rm cc}$).
The top panel is the 
total cloud mass normalized by the initial cloud mass ($M_c / M_{c,0}$),
the middle panel is the 
mass-weighted average velocity of the cloud along the wind direction normalized by the wind velocity ($v_{x,c} / v_w$),
and the bottomn panel is the cloud growth time
$t_{\rm grow}\equiv M_c / \dot{M_c}$
normalized by $t_{\rm cc}$.
The cloud gradually grows by accreting the cooled wind material when radiative cooling in the cloud-wind interface is efficient ($n^\prime_{\rm c} > 1$).
The cloud accretes about ten times its original mass from the wind 
by the end of the simulation ($t/t_{\rm cc} = 24$).
On the other hand, the cloud is destroyed by fluid instability in a few $t_{\rm cc}$ when cooling is inefficient ($n^\prime_{\rm c} < 1$).
The initial cloud material is accelerated as it is pushed by the fast-moving wind, and the acceleration history is similar in all cases, approaching the wind velocity at $t/t_{\rm cc} \approx 10$, which we shall refer to as the beginning of \textit{entrainment}.
Note that this entrainment time is of the order of the drag time $t_{\rm drag}\sim \chi^{1/2} t_{\rm cc}$ for the $T_{\rm w}=10^6\,$K runs, but is significantly shorter for the higher overdensity runs where $T_{\rm w}=10^7\,$K. 
Here, the acceleration comes predominantly from momentum transfer due to the cold gas growth \citep[cf.][]{Gronke2020,2021ApJ...911...68T}, i.e., occurs on a timescale of $t_{\rm grow}$, which for all of our cloud-surviving runs is $\sim 5-10 t_{\rm cc}$\footnote{While nominally the growth time is $t_{\rm growth}\equiv M_{\rm c}/\dot M_{\rm c}\sim \chi V_{\rm c} / (A v_{\rm mix})$ \citep{Gronke2020}, the cold gas area-to-volume ratio is significantly larger in our high-$\chi$ runs, thus, leading to a relatively short growth, and thus, entrainment time.}.

Our results confirm the well-established cloud survival criterion, which is reassuring.
We shall focus on the more interesting simulations where $n^\prime_{\rm c} > 1$ for the remainder of this work.

\begin{figure*}
	\centering
	\includegraphics[width=0.99\linewidth]{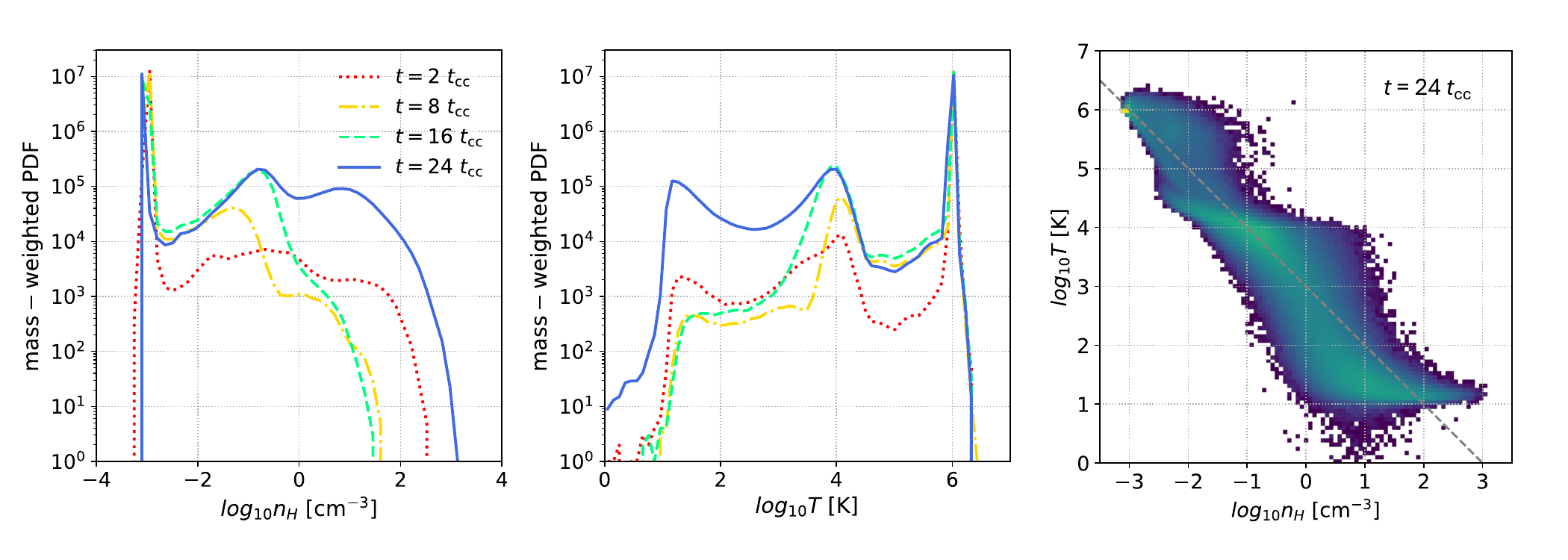}
	\caption{
		The probability distribution functions of the hydrogen number density (left) and temperature (middle)
		for the simulation with $T_{\rm w} = 10^6$~K and $n^\prime_c = 3$ (or $n_c = 0.1~{\rm cm^{-3}}$) 
		at $t/t_{\rm cc} = $~2, 8, 16, and 24.
		The right panel shows the phase diagram at $t/t_{\rm cc} = 24$,
		with the grey dashed line indicating constant pressure.
		The cold phase of the cloud is significantly denser than the initial cloud density due to compression.
	}
	\label{fig:pdpdf}
\end{figure*}

\subsection{Formation of the Two-phase Clouds}

Fig.~\ref{fig:maps_n_T} shows snapshots of the hydrogen number density (panel a) and gas temperature (panel b) slices for the simulation with $n^\prime_{\rm c} = 3$ and $T_{\rm w} = 10^6$~K at $t / t_{\rm cc} = $ 0, 2, 8, 16, and 24 from top to bottom.
As the simulation progresses, the cloud grows by accreting cooled material from the wind, eventually forming an elongated tail, consistent with previous studies.
However, unlike simulations that exclude cooling below $10^4$~K, the cold gas ($T \lesssim 10^2$ K) develops within the warm cloud ($T\sim 10^4$ K) at late times due to fine-structure metal line cooling.
The complex structure of the cold gas is most evident in the zoom-in panels (right) at $t = 24 ~t_{\rm cc}$, with densities two orders of magnitude higher than the surrounding warm gas. 

To further quantify the multiphase structure of the cloud, Fig.~\ref{fig:pdpdf} shows the probability distribution functions (PDFs) of the hydrogen number density (left) and temperature (middle) at $t / t_{\rm cc} = $~2, 8, 16, and 24.
At $t = 2 ~t_{\rm cc}$, a significant amount of cold gas has already developed, resulting in a bimodal temperature PDF with peaks at $T \sim 10^4$~K and $T \sim 20$~K.
This early cold gas is primarily localized at the cloud's leading edge, where the material is shocked and compressed by the wind.
However, this phase is short-lived, rapidly disrupted by fluid instabilities before accelerating to high velocities.
As the cloud evolves, an elongated tail forms, initially dominated by warm gas.
Once the cloud becomes fully entrained, a secondary phase of cold, dense gas emerges -- this time within the tail, covered by the surrounding warm, diffuse medium.
This is in good agreement with \citet{Chen2024} who showed that the cold gas forms after entrainment. 
The temperature PDF becomes bimodal again at late times ($t = 24 ~t_{\rm cc}$) due to the continuous development of the cold gas, with densities reaching $n \sim 100~{\rm cm^{-3}}$, approximately two orders of magnitude higher than the initial cloud density.
The phase diagram (right panel) at $t = 24 ~t_{\rm cc}$ indicates that the three phases are in approximate pressure equilibrium, though the cold phase remains slightly under-pressured.
This distribution closely resembles the classical three-phase interstellar medium (ISM) and provides an ideal environment for both dust growth and H$_2$ formation, as discussed in the following sections.

\begin{figure*}
	\centering
	\includegraphics[width=1\linewidth]{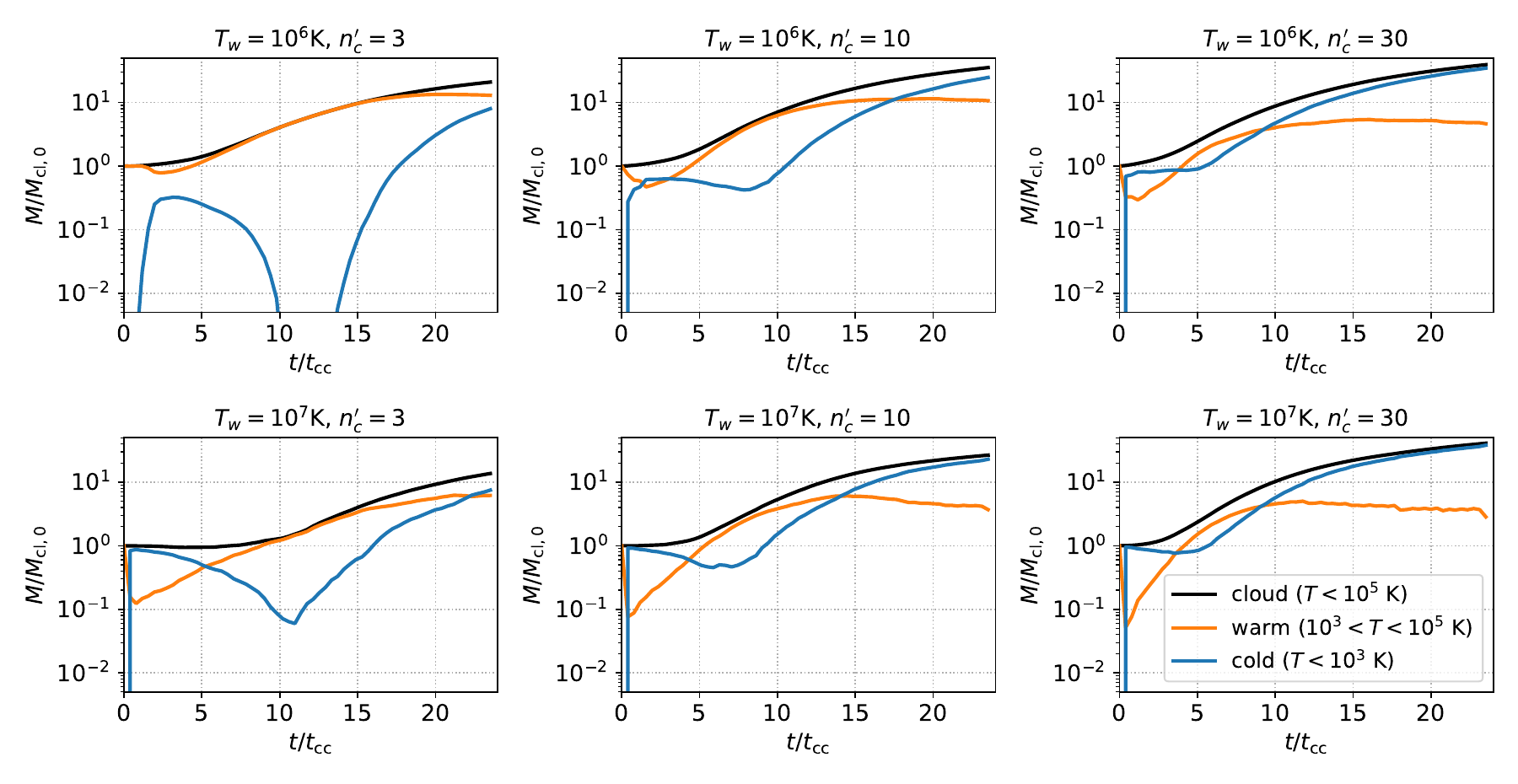}
	\caption{
		Time evolution of the 
		total mass of the cloud ($T < 10^5$ K, black),
		which is further divided into
		the warm gas ($10^3 < T < 10^5$ K, orange) and cold gas ($T < 10^3$ K, blue),
		for simulations with different setups:
		top panels are for wind temperature $T_{\rm w} = 10^6$~K
		while
		bottom panels are for $T_{\rm w} = 10^7$~K.
		The initial cloud density varies from $n_{\rm c}$ = 3, 10, and 30 $n_{\rm crit}$ from left to right.
		The cloud becomes dominated by the cold gas for $n_{\rm c} \geq 10~n_{\rm crit}$.
	}
	\label{fig:warmcold_time}
\end{figure*}

Fig.~\ref{fig:warmcold_time} shows the total mass of the cloud (black), warm gas (orange), and cold gas (blue) as a function of time for the six of our simulations where clouds survive and grow.
In all cases, the total cloud mass (the sum of the cold and warm phases) increases over time, as expected for simulations in the cooling-dominated regime ($n^\prime_{\rm c} > 1$).
By the end of the simulations, cloud masses have increased by factors of 10 to 40, with higher initial cloud densities resulting in slightly more rapid accretion.
As shown in Fig.~\ref{fig:timescale},
the isobaric cooling time at $T \lesssim 10^4$~K is $t_{\rm cool} \sim 10 P_3^{-1}~{\rm Myr}$ where $P_3 = P/(10^3{\rm K~cm^{-3}})$.
Since $P_3 \geq 1$ in all our simulation runs (cf. Table\ref{tab:params}),
gas cools below $10^4$~K rapidly within the first 10 Myr $\approx 2~t_{\rm cc}$ or less.
While cold gas forms almost immediately at the onset of the simulations due to compression and cooling at the cloud front, this material is quickly disrupted by fluid instabilities.
A second, more sustained growth phase follows as cold gas forms within the developing tail.
Higher initial cloud densities lead to clouds that are increasingly dominated by the cold phase.

\subsection{Dust Evolution}

\begin{figure*}
	\centering
	\includegraphics[width=1\linewidth]{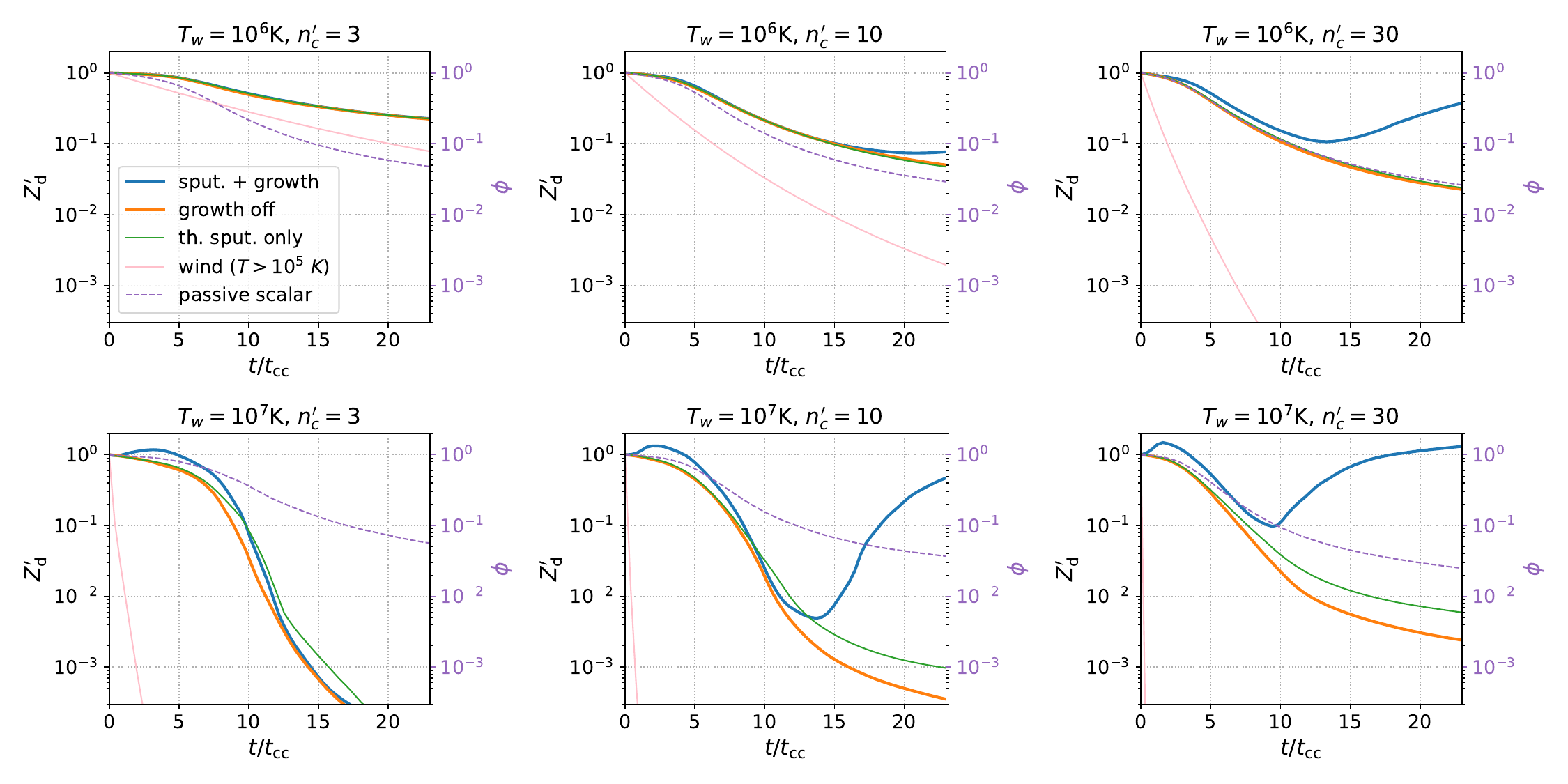}
	\caption{
		Time evolution of the dust-to-gas ratio (DGR) normalized to the Milky Way value ($Z_{\rm d}^\prime \equiv Z_{\rm d} / Z_{\rm d,MW}$) in the cloud for three different dust models:
        (i)
        the fiducial dust model (sputtering + growth, solid blue),
        (ii)
		a model with dust growth switched off (solid orange),
		and 
        (iii)
        a model with nonthermal sputtering switched off (solid green).
		The passive scalar abundance 
		(from the fiducial dust model, though it is very similar in all three dust models) 
		is shown in dashed purple lines.
		The DGR in the wind 
		(also from the fiducial dust model)
		is shown in solid pink lines.
		For $T_{\rm w} = 10^6$~K cases, dust sputtering occurs almost exclusively in the wind, 
		and thus the DGR in the clouds is diluted via mixing with the dust-depleted wind.
		For $T_{\rm w} = 10^7$~K cases, the DGR declines much faster than the passive scalar, 
		indicating additional sputtering during the mixing process.
		Dust growth significantly enhances the DGR at late times for $n_{\rm c}^\prime \gtrsim 10$.
		Nonthermal sputtering has a negligible effect compared to thermal sputtering.
	}
    \vspace{2mm}
	\label{fig:fig_dgr_sputter_vs_growth}
\end{figure*}

Fig.~\ref{fig:fig_dgr_sputter_vs_growth} shows the time evolution of the dust-to-gas ratio (DGR) in the cloud
in three different dust evolution models: the fiducial model with dust growth and sputtering (solid blue), the sputtering-only model with dust growth switched off (solid orange), and the model with nonthermal sputtering switched off (dashed green).
In addition, we show the passive scalar field ($\phi$, dashed purple), which traces the initial cloud material, only for the fiducial dust model.
We do not show $\phi$ for the other models as its evolution is nearly identical across all three runs.
This is expected since the DGR only marginally affects the (thermo)dynamics of the clouds through photoelectric heating and H$_2$ rovibrational cooling, both of which are not expected to be efficient in our setup.
For the same reason, the DGR in the wind is plotted only for the fiducial model (dashed red).

\begin{figure}
	\centering
	\includegraphics[width=1\linewidth]{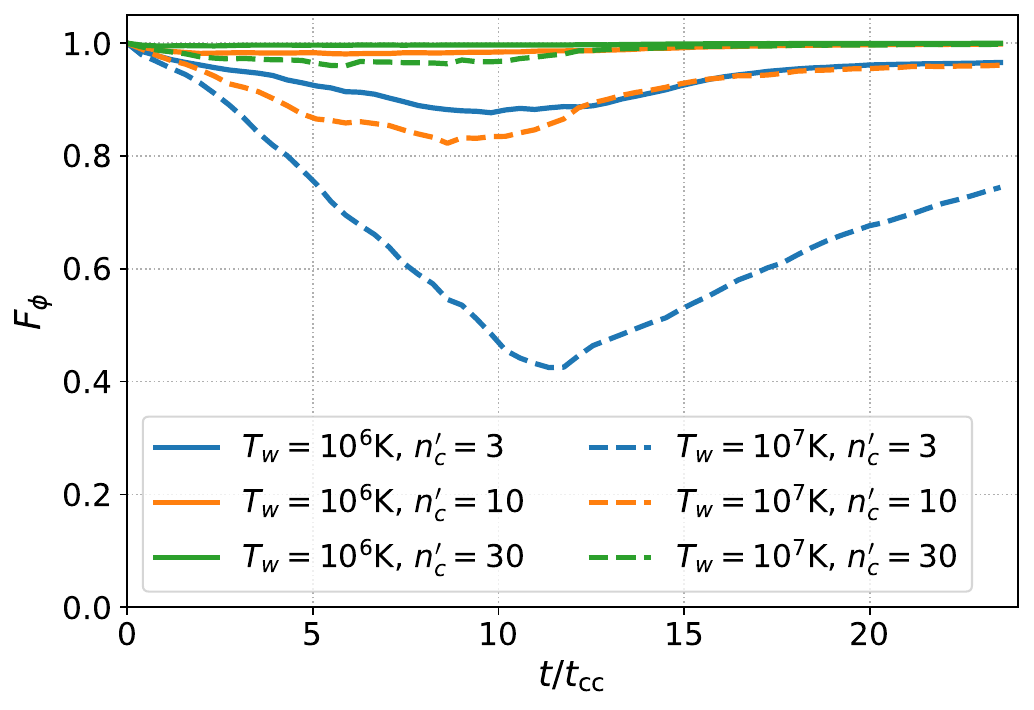}
	\caption{
        Mass fraction of the initial cloud material that is currently in the cloud.
        The majority of the initial cloud material remains in the cloud throughout the simulation.
	}
	\label{fig:Mphi_time}
\end{figure}

\begin{figure*}
	\centering
	\includegraphics[width=1.0\linewidth]{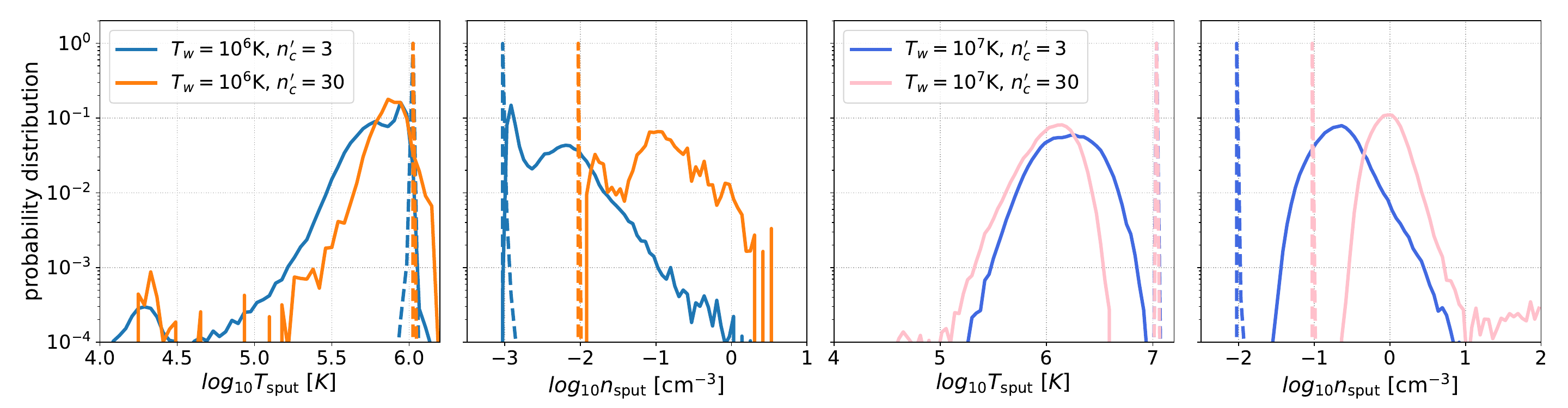}
	\caption{
        Probability density distribution of the sputtering temperature $T_{\rm sput}$ and sputtering density $n_{\rm sput}$ (see text for definitions) for gas originally in the cloud (solid) and in the wind (dashed).
        For the $T_{\rm w} = 10^6$~K case, 
        thermal sputtering occurs almost exclusively in the wind.
        In contrast, for the $T_{\rm w} = 10^7$~K case, sputtering occurs both in the wind and in the intermediate-temperature mixing layers. 
	}
	\label{fig:sput_T_n}
\end{figure*}

\subsubsection{Dust Destruction via Sputtering}

We first examine the sputtering-only model (solid orange curve in Fig.~\ref{fig:fig_dgr_sputter_vs_growth}).
For $T_{\rm w} = 10^6$~K and $n^\prime_{\rm c}$ = 30, the dust and passive scalar ($Z^\prime_{\rm d}$ and $\phi$, respectively) follow an almost identical time evolution. 
This indicates that dust destruction takes place almost exclusively in the hot wind where sputtering is so efficient that the wind becomes essentially dust-free immediately after the simulation begins.
Conversely, dust initially residing in the cloud remains shielded from the hot gas and thus stays intact throughout the simulation.
In this regime, the evolution of $Z^\prime_{\rm d}$ primarily reflects the mixing between the dusty cloud material and the dust-free wind.
The cloud's DGR is gradually diluted as it accretes dust-depleted wind material, driven by mixing and subsequent radiative cooling.

For the cases of $T_{\rm w} = 10^6$~K and $n^\prime_{\rm c}$ = 3 and 10, 
$Z^\prime_{\rm d}$ remains slightly higher than $\phi$.
This occurs because the wind is not entirely dust-free, as sputtering in the hot gas is less efficient at these lower densities. 
Consequently, the dilution effect is less pronounced than in the high-density case.

The evolution of the DGR behaves very differently in the $T_{\rm w} = 10^7$~K simulations (lower panels of Fig.~\ref{fig:fig_dgr_sputter_vs_growth}) from the $10^6$~K wind.
First, due to the higher temperatures and densities (clouds have to be significantly denser to survive), dust in the wind is destroyed almost instantaneously at the start of the simulation.
As such, the wind remains effectively dust-free in all three different cases of $n^\prime_{\rm c}$.
More importantly, the cloud's DGR declines significantly faster than the passive scalar.
This suggests that, besides simple dilution, dust originally residing in the cloud at $t = 0$ undergoes substantial sputtering as the hot wind accretes onto the cloud.

To understand how this additional sputtering occurs,
Fig.~\ref{fig:Mphi_time} shows the mass fraction of the initial cloud material that is currently in the cloud ($F_\phi$) as a function of time.
In most cases, the majority ($>80\%$) of the initial cloud material remains in the cloud throughout the simulation,
with the exception of the case $T_{\rm w} = 10^7$~K and $n^\prime_{\rm c} = 3$,
where $F_\phi$ drops to $\sim 40\%$ at its minimum.
If the dust originally in the cloud (the ``cloud dust'') were destroyed because it entered the hot wind,
the ratio of $\phi$ to $Z^\prime_{\rm d}$ would roughly follow $F_\phi$ in the sputtering-only runs.
However, 
this ratio is significantly lower as shown in Fig.~\ref{fig:fig_dgr_sputter_vs_growth}.
Therefore, sputtering of the cloud dust is not controlled by how much of it can escape the cloud phase and become part of the wind.

Instead, sputtering of the cloud dust occurs primarily in the cloud-wind interface at intermediate temperatures.
To demonstrate this, we define the sputtering temperature ($T_{\rm sput}$) and sputtering density ($n_{\rm sput}$) as, respectively, the average gas temperature and density over the simulation time weighted by the sputtered dust mass tracked in every timestep throughout the simulations.
Fig.~\ref{fig:sput_T_n} shows the probability density distributions of $T_{\rm sput}$ and $n_{\rm sput}$ for gas originally in the cloud (solid) and in the wind (dashed).
In the $T_{\rm w} = 10^6$~K wind, sputtering occurs almost exclusively in the wind.
The majority of the cloud dust survives, with only a small fraction destroyed as it enters the wind phase.
In contrast, in the $T_{\rm w} = 10^7$~K wind, a significant amount of cloud dust is sputtered within the cloud-wind interface (the mixing layer) at intermediate temperatures.
This difference arises because at $T_{\rm w} = 10^7$~K, the mixing layer temperature is high enough to drive efficient sputtering and destroy the cloud dust ($T_{\rm mix} = \sqrt{T_{\rm w} T_c} \sim 3.2\times 10^5$~K following \citealp{BegelmanFabian1990} but also see, e.g., \citealp{Tan2021,2021MNRAS.508L..37T,SharmaTRML2025} for detailed mixing-layer simulations).
Note that the thermal sputtering rate decreases sharply with temperature in this regime,
and the interface temperature is only $\sim 10^5$~K for the $T_{\rm w} = 10^6$~K wind,
corresponding to a sputtering rate approximately fifty times lower.
As such, sputtering is essentially shut off in the interface, and the DGR evolution is almost exclusively dictated by the dilution effect.


\subsubsection{Re-formation of Dust}

\begin{figure}
	\centering
	\includegraphics[width=1\linewidth]{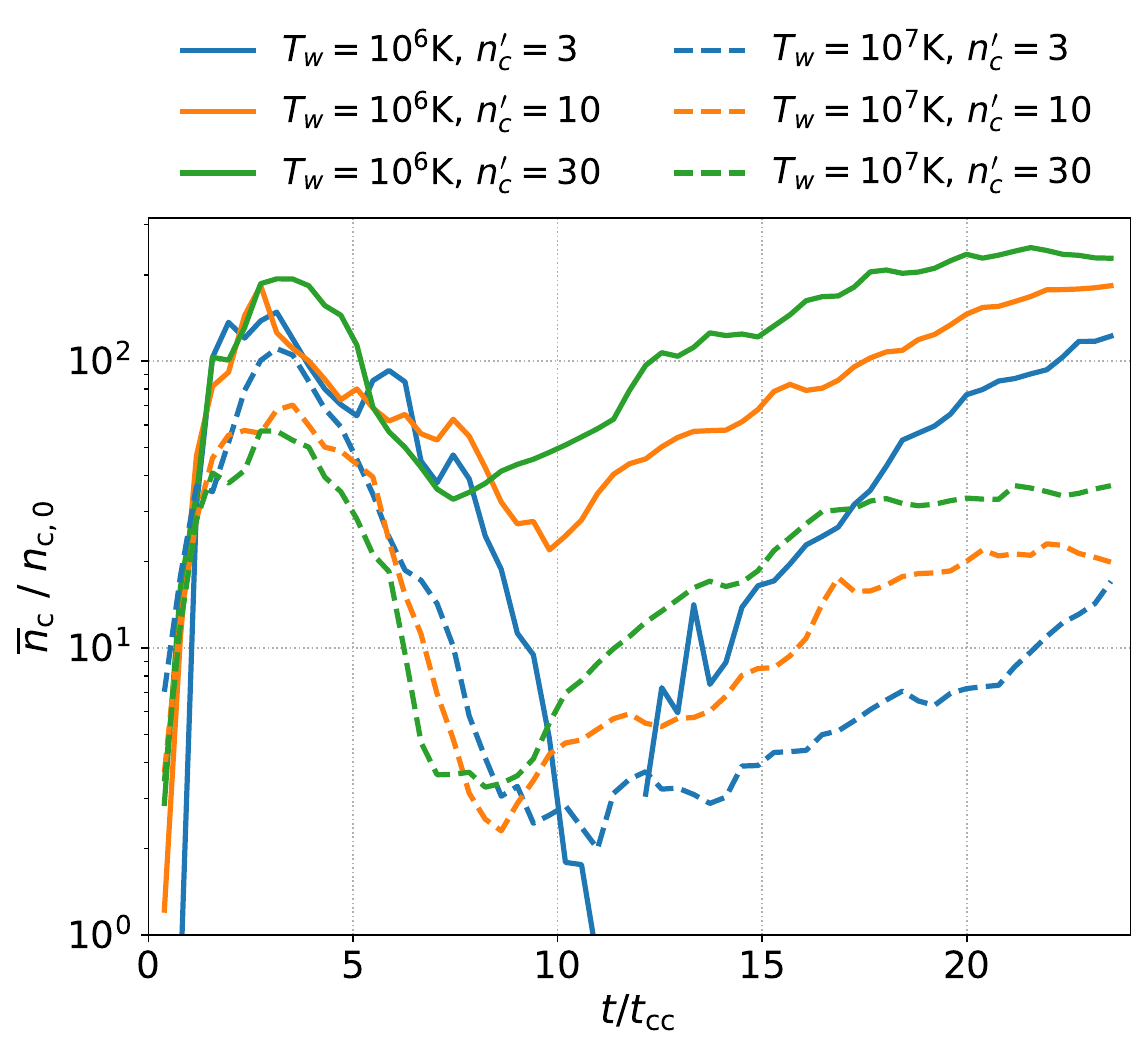}
	\caption{
		Time evolution of the
		mass-weighted density of the cold gas $\bar{n}_{\rm cold}$ 
		normalized by the initial cloud density $n_{\rm c,0}$.
        The elevated density of the cold gas significantly enhances dust growth and H$_2$ formation.
	}
	\label{fig:nH_evol}
\end{figure}

In the sputtering-only simulations, dust can only be destroyed as no formation mechanism is included.
When dust growth is included in our fiducial model (solid blue curves in Fig.~\ref{fig:fig_dgr_sputter_vs_growth}), the evolution of dust becomes a competition between destruction and formation, which happen in different gas phases.
Sputtering occurs mostly in the hot wind, while dust growth is restricted to the cold phase of the cloud, where densities are significantly higher than the initial cloud density.
The dust growth rate of the cloud can be obtained by
$\dot{M}_{\rm d} = \sum_i \dot{m}_{{\rm d},i}
\sim 0.67~{\rm Gyr^{-1}}\sum_i (m_{\rm d} n_0 T_2^{0.5})_i$ 
where $n_0 = n/{\rm cm^{-3}}$, $T_2 = T/{\rm 100 ~K}$, and the summation is over the cold gas particles.
Assuming that the cold gas has a characteristic temperature of 30 K 
and that the DGR in the cloud is well-mixed,
the dust growth time of the cloud can be expressed as 
$t_{\rm grow} = M_{\rm d} / \dot{M}_{\rm d} \sim 2.7~{\rm Gyr} (F_{\rm cold} \bar{n}_{\rm cold})^{-1}$ where  
$F_{\rm cold}$ is the cold gas fraction and 
$\bar{n}_{\rm cold}$ is the mass-weighted mean density of the cold gas.
Comparing it to the cloud-crushing time, 
we obtain $t_{\rm grow} / t_{\rm cc} \sim 1.8\times 10^4 (\bar{n}_{\rm cold}/n_{\rm crit})^{-1}$ for $T_{\rm w} = 10^6$~K and $t_{\rm grow} / t_{\rm cc} \sim 1.8\times 10^2 (\bar{n}_{\rm cold}/n_{\rm crit})^{-1}$ for $T_{\rm w} = 10^7$~K, respectively. 
Therefore, the critical parameter governing the dust growth rate is the ratio $\bar{n}_{\rm cold}/n_{\rm crit} = (\bar{n}_{\rm cold}/n_{\rm c}) n^\prime_{\rm c}$.

\begin{figure*}
	\centering
	\includegraphics[width=0.9\linewidth]{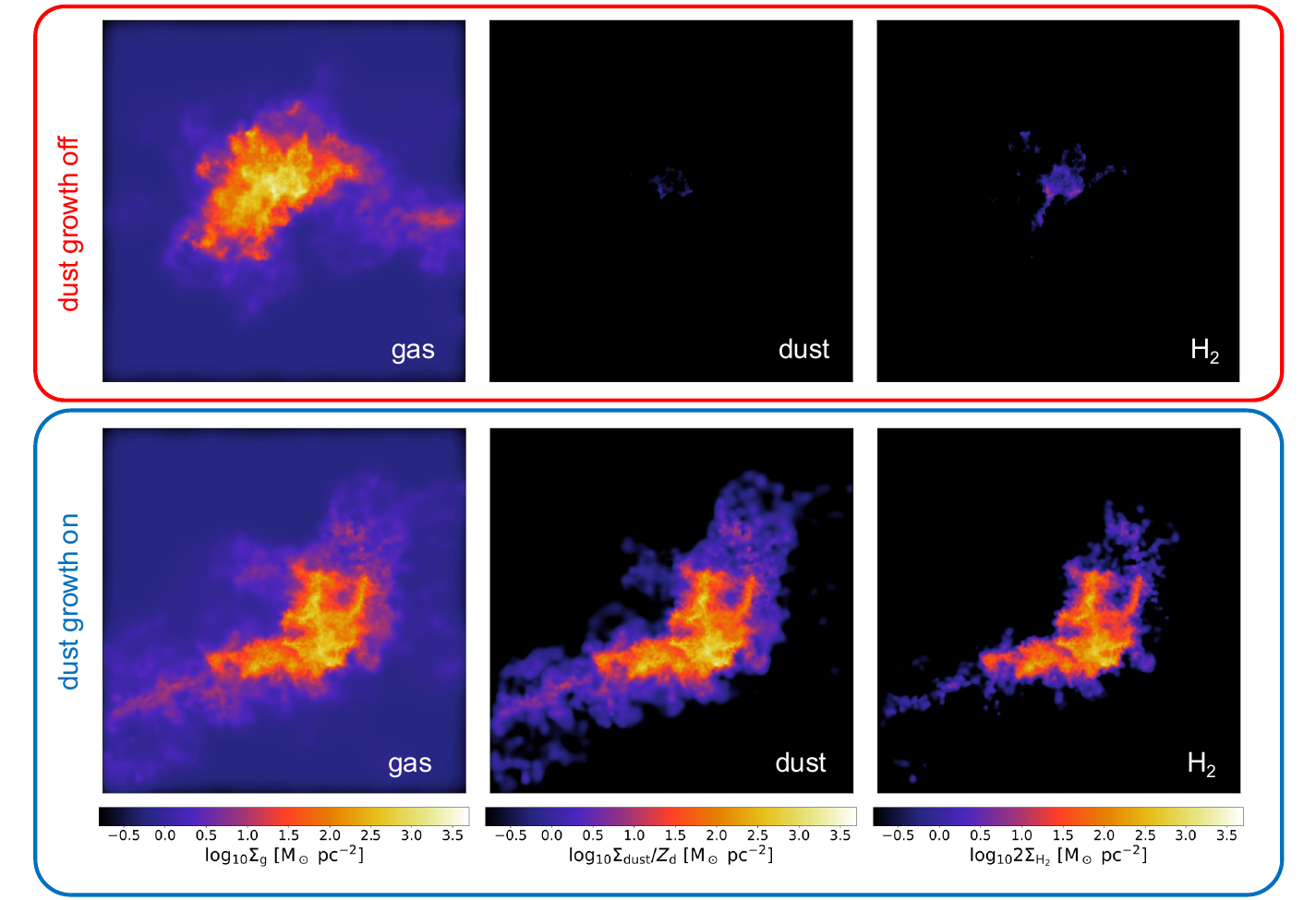}
	\caption{
    Surface density images of the total gas (left), dust (middle), and H$_2$ (right) focusing on one of the cloud fragments at $t = 24~t_{\rm cc}$ for the case of $T_{\rm w} = 10^7$~K and $n^\prime_{\rm c}$ = 10. The upper and lower panels show the models without and with dust growth, respectively.
	}
    \vspace{2mm}
	\label{fig:mapsdustgrowth}
\end{figure*}

In Fig.~\ref{fig:nH_evol}, we show the time evolution of $\bar{n}_{\rm cold}/n_{\rm c}$ for all six simulations.
Initial shock compression triggers a transient density spike, followed by a decline as the primary cloud material is disrupted.
Once the cloud is entrained, $\bar{n}_{\rm cold}/n_{\rm c}$ increases again as cold gas reforms in the tail.
The degree of density enhancement depends on both $n_{\rm c}$ and $T_{\rm w}$.
Higher initial cloud densities promote more significant enhancement through increased cold gas formation.
Additionally, we find that the density enhancement is inversely related to $T_{\rm w}$ as the cold gas in the $T_{\rm w} = 10^7$~K cases is more severely under-pressurized.

For $T_{\rm w} = 10^6$~K, adopting $\bar{n}_{\rm cold} / n_{\rm c} \sim 200$, we obtain $t_{\rm grow} / t_{\rm cc} \sim 90 / n^\prime_{\rm c}$, which corresponds to 30, 9, and 3
for our runs of $n^\prime_{\rm c} = $ 3, 10, and 30, respectively.
Therefore, dust growth is only significant in the case of $n^\prime_{\rm c} = 30$ (see Fig.~\ref{fig:fig_dgr_sputter_vs_growth}).
On the other hand, for $T_{\rm w} = 10^7$~K, adopting $\bar{n}_{\rm cold} / n_{\rm c} \sim 20$, we obtain $t_{\rm grow} / t_{\rm cc} \sim 0.9 / n^\prime_{\rm c}$, which corresponds to 3, 0.9, and 0.3 for our runs of $n^\prime_{\rm c} = $ 3, 10, and 30, respectively.
While there is sufficient time for dust to form in all cases, sputtering is also much more efficient here.
As such, sputtering of the original cloud dust leads to a much faster decline of $Z^\prime_{\rm d}$ than simple dilution from the dust-free wind material.
In the low-density case of $n^\prime_{\rm c} = $ 3, although there is enough time for dust to grow ($\sim 3 ~t_{\rm cc}$), sputtering outcompetes dust growth and thus $Z^\prime_{\rm d}$ decreases monotonically, leading to an essentially dust-free cold phase.
Dust growth only dominates over sputtering at higher initial densities ($n^\prime_{\rm c} = $ 10 and 30), where $Z^\prime_{\rm d}$ eventually recovers its initial value, orders of magnitude higher than the DGR values in the sputtering-only runs.

Nonthermal sputtering plays a negligible role in the DGR evolution (green dashed lines).
This contrasts with supernova remnants (SNRs), where both thermal and nonthermal sputtering are comparably important \citep{Slavin2015, Hu2019a}.
The distinction can be attributed to the Mach number of the flow: our cloud is transonic ($\mathcal{M}\sim 1$), while SNRs involve blastwaves and highly supersonic flows that lead to high grain-gas drift velocities required for non-thermal processes.
Furthermore, the destruction of cloud dust via nonthermal sputtering requires the cloud to be strongly accelerated to generate high dust-gas relative velocities, which only occurs in the initial shock phase.
Once the cloud is entrained, there is no mechanism to kinematically decouple dust and gas, and thus nonthermal sputtering becomes inactive. 
In contrast, thermal sputtering only requires sufficiently high temperatures in the cloud-wind interface, which remains active even after entrainment.

\vspace{-1mm}

\begin{figure*}
	\centering
	\includegraphics[width=1\linewidth]{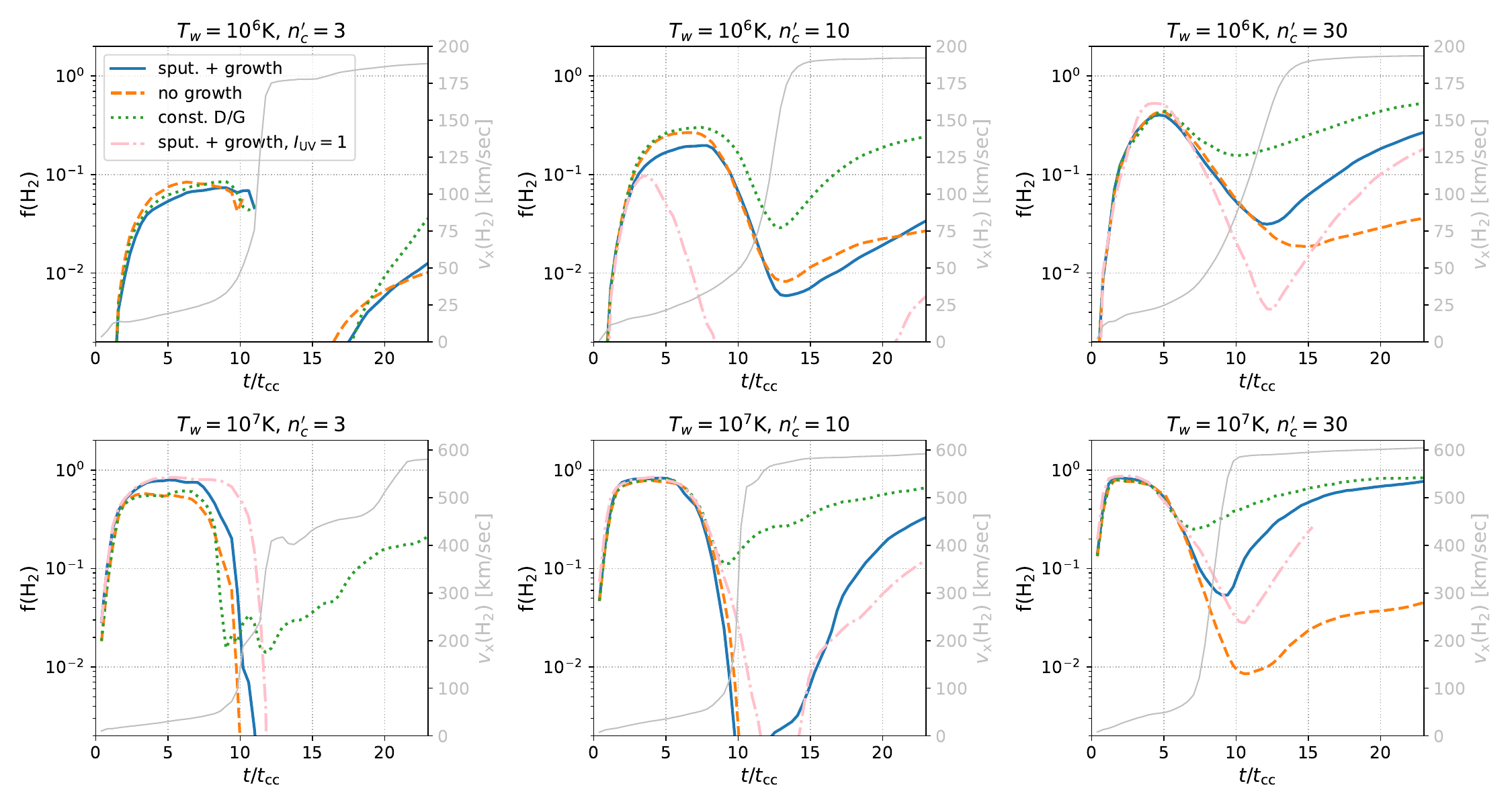}
	\caption{
		Time evolution of the H$_2$ mass fraction of the clouds ($T < 10^5$ K) in the fiducial model (with dust growth and sputtering, solid blue lines), sputtering-only model (dashed orange), model with no dust evolution at all (dotted green), and fiducial dust model with a radiation field $I_{\rm UV} = 1$ (dash-dotted pink).
        The H$_2$-mass-weighted velocity in the wind direction is shown in solid grey lines with scales indicated by the secondary $y$-axis on the left. 
		Sputtering significantly suppresses H$_2$ formation, while dust growth can counteract sputtering for dense clouds at late times.
	}
	\label{fig:fig_H2_sput_vs_growth_time_evol}
\vspace{2mm}
\end{figure*}

\vspace{2mm}
\subsection{Formation of Molecular Gas}

The formation of H$_2$ depends crucially on the presence of dust.
The upper panels of Fig.~\ref{fig:mapsdustgrowth} show the surface density images of the total gas (left), dust (middle), and H$_2$ (right), zoomed in on a cloud fragment at $t = 24~t_{\rm cc}$
for the case of $T_{\rm w} = 10^7$~K and $n^\prime_{\rm c}$ = 10 with dust growth disabled.
In this case, the cloud is nearly dust-depleted due to efficient sputtering and is dominated by atomic hydrogen despite a low-temperature, high-density environment that is otherwise favorable for H$_2$ formation.
Conversely, when dust growth is switched on (lower panels), the re-formation of dust at late times provides the necessary catalyst for H$_2$ formation, and the cloud becomes significantly molecular. 
While H$_2$ is largely confined to the cold, dense inner core of the cloud, dust is more broadly distributed, extending into the warm, diffuse outer envelope.
The difference arises from mixing between the warm and cold phases: dust forms in the cold gas and subsequently diffuses into the warm gas.
In contrast, while H$_2$ also forms in the cold gas, it is rapidly dissociated by ambient radiation upon entering the warm envelope due to a lack of shielding.

To quantify the impact of dust evolution on H$_2$ formation, Fig.~\ref{fig:fig_H2_sput_vs_growth_time_evol} shows the time evolution of the H$_2$ mass fraction in the cloud in three different dust models: the fiducial model (sputtering and growth, solid blue), the sputtering-only model (growth disabled, dashed orange), and the constant DGR model (no dust evolution, dotted green).
The constant DGR model essentially represents an upper limit for H$_2$ formation as we adopt the Milky Way dust-to-metal ratio, which is close to unity.
Given that the H$_2$ formation time at $Z_{\rm d}^\prime = 1$ is comparable to the dust growth time, the cloud has sufficient time to become significantly molecular by the end of the simulations in all cases except for $T_{\rm w} = 10^6$~K and $n^\prime_{\rm c}$ = 3 (where $t_{\rm H_2} / t_{\rm cc} \sim 30$).

In contrast, the sputtering-only model leads to HI-dominated clouds in all cases.
For $10^7$~K wind, the DGR is depleted by orders of magnitude, effectively shutting off H$_2$ formation  -- thus, severely overestimating $f(H_2)$ in the constant DGR cases.
For $10^6$~K wind, although sputtering is relatively mild and driven primarily by dilution, H$_2$ formation is also slow because of lower cloud densities. 
Therefore, a factor of ten reduction in $Z^\prime_{\rm d}$ is sufficient to restrict the H$_2$ fraction to only a few percent.

The inclusion of dust growth in the fiducial model fundamentally changes the results.
The late-time dust reformation significantly facilitates H$_2$ formation, leading to high molecular fractions at $n^\prime_{\rm c}$ = 30 in the $10^6$~K wind and $n^\prime_{\rm c} \geq 10$ in the  $10^7$~K wind.
At lower initial cloud densities, the cloud remains atomic due to persistent dust depletion.
These results highlight that H$_2$ formation is not only a function of gas density and temperature but is largely regulated by the survival and growth of dust.

Note that while the clouds also exhibit an H$_2$-dominated phase at early times ($t / t_{\rm cc} \sim 5$), this phase is transient and travels at low velocities,
and thus should not be classified as a true ``molecular wind.''
This is demonstrated by the H$_2$ mass-weighted velocity (solid grey lines in Fig.~\ref{fig:fig_H2_sput_vs_growth_time_evol}), which remains significantly lower than the ambient wind speed during this stage.
In contrast, the H$_2$ gas that forms at late times travels at nearly the wind velocity once the clouds have become fully entrained. 
It is this entrained molecular gas, rather than the early transient material, that corresponds to the high-velocity spectrum components typically observed in galactic outflows.

Finally, we investigate the effect of the FUV radiation field.
Our adopted value of $G_0 = 3.2\times 10^{-3}$ is representative of extraplanar gas far from the galactic disk.
Conversely, if gas is still close to the disk, the radiation field should approach standard ISM values.
We therefore run an additional set of simulations with $G_0 = 1.7$ (the Draine field) with the fiducial dust model (pink dash-dotted lines in Fig.~\ref{fig:fig_H2_sput_vs_growth_time_evol}). 
A stronger radiation field penetrates deeper into the cloud via photodissociation, effectively confining H$_2$ to the most well-shielded, cold, and dense regions.
However, since the majority of the molecular mass resides in the dense phase, which remains well-shielded even at $G_0 = 1.7$, the H$_2$ fraction is only modestly suppressed.
The most pronounced suppression occurs in the $n^\prime_{\rm c}$ = 10, $T_{\rm w} = 10^6$~K case, as the H$_2$ gas here is of the lowest densities and column densities and is thus most sensitive to the adopted $G_0$.


\section{Discussion}
\label{sec:discussion}

\subsection{Comparison with Previous Work}
While a substantial body of work has examined classical ``cloud-crushing'' setups (e.g., \citealp{Scannapieco2015,Schneider2016,Gronke2018}; see also \S~\ref{sec:intro}), which informs the design of our study, we focus here specifically on previous work addressing dust and molecular evolution.

\citet{Girichidis2021} investigate \textit{in situ} H$_2$ formation in winds using cloud-crushing simulations coupled with a chemistry network similar to ours. 
They include magnetic fields and self-gravity, which we neglect in this work.
They show that the cloud develops into a two-phase medium when cooling below $10^4$~K is allowed,
and that the cloud density is the key factor dictating the evolution, both of which are consistent with our results.
A limitation of their work is the assumption of a constant DGR, which, as we have shown, significantly overestimates H$_2$ formation.
Their H$_2$ fractions should thus be regarded as upper limits.
Furthermore, their simulations are only run for a few cloud-crushing times, focusing on the early evolution of the cloud before entrainment.
Consequently, the molecular gas has not yet accelerated to high velocities, and the long-term evolution remains unclear.
Our results suggest that the majority of this early-stage molecular gas could be destroyed by $t/t_{\rm cc} \gtrsim 10$.

\citet{Farber2022} study thermal sputtering in cloud-crushing simulations with an initial cloud temperature of $10^3$ K and extend the cooling function down to 300 K.
They generalize the survival criterion \citep{Gronke2018} to lower cloud temperatures and find that two regimes of `only warm' or `cold gas survival' exist. Here, our runs fall in the latter regime.
\citet{Farber2022} do not use a sophisticated chemical network or explicit dust modeling but instead include tracer particles to track the temperature evolution of gas, and they define a threshold temperature $T_{\rm dest}$ above which dust is assumed to be destroyed by sputtering.
They treat $T_{\rm dest}$ as a free parameter and explore its impact on dust survival.
They find that cold gas and dust can survive in the cloud for large clouds.
In their $T_{\rm dest} = 10^5$~K model, which is arguably more realistic, they find that nearly all dust survives in their 100 pc cloud, in qualitative agreement with our results.

The cloud-crushing simulations run by \citet{Chen2024} are the most comparable to ours.
They include dust growth and thermal sputtering and allow cooling below $10^4$ K with a piece-wise power-law cooling function extended down to 10 K.
They also run $n^\prime_{\rm c} = $ 3, 10, and 30 in a $10^6$ K wind.
Contrary to our results, however, they find that dust growth is inefficient in all cases.
This discrepancy might be caused by their adopted grain size of 0.1 $\mu m$, which is approximately three times our MRN-average value, leading to a slower dust growth rate.
In other words, dust growth might start to become significant at $n^\prime_{\rm c} \gtrsim 100$ in their setup. 
The main caveat of their work is that they do not model H$_2$ formation explicitly.
Indeed, they investigate cold gas formation and carefully use the term ``molecular temperature'' to refer to gas temperature low enough for H$_2$ formation, while remaining agnostic to the chemical composition of the gas.
As we have shown, the existence of a cold phase does not guarantee H$_2$ formation.
Indeed, without dust growth, the DGR of the cloud would be too low for H$_2$ to form in all cases, either due to dilution or cloud dust sputtering.
In this case, the cloud would still be mostly atomic despite the low-temperature, high-density conditions.
Finally, \citet{Chen2024} account for shielding against the cosmic UV background via a temperature floor at $10^4$~K in regions with column densities below $10^{19}~{\rm cm^{-2}}$.
They show that photoionization heating only modestly affects the long-term evolution of different temperature phases, as clouds capable of surviving fluid instabilities are also sufficiently self-shielded.

\citet{Richie2024} run cloud-crushing simulations that include thermal sputtering and a cooling floor at $10^4$ K.
Their main focus is the survival of dust, with no discussion on H$_2$ formation.
They find that a significant amount of dust survives in the cloud.
In addition, they adopt two different wind temperatures, $3\times 10^6$ K and $3\times 10^7$ K, and find more significant sputtering of the cloud dust in their high-temperature wind, consistent with our results.
However, the absence of a cold phase in the cloud means that dust growth cannot operate, which they indeed neglect.
Without dust growth, the DGR is reduced by an order of magnitude in the tail, in agreement with our sputtering-only runs.
As we have shown, including dust growth and enabling cooling below $10^4$ K (such that dust growth can occur in the cold phase) is critical to keep the cloud's DGR high enough for H$_2$ formation.

\vspace{10mm}
\subsection{The Origin of Multiphase Galactic Winds}

Large quantities of cold ($\sim 10^{4}\,\mathrm{K}$) gas are ubiquitously detected at high velocities in galactic winds \citep{Rupke2018,Veilleux2020}. Since the primary agents driving these winds—such as supernova feedback—initially accelerate hot gas, these observations imply that galactic winds are intrinsically multiphase. They also raise the long-standing question of how cold gas can be accelerated to high velocities before being disrupted or destroyed \citep{Zhang2015a}.

One proposed solution is that cold gas forms \emph{in situ} within the wind. In this picture, the adiabatic expansion of the outflow leads to a decrease in wind temperature beyond a critical radius, enabling efficient radiative cooling and the condensation of fast-moving cold gas \citep{Wang1995,Thompson2015}. An alternative scenario is that pre-existing cold gas is entrained and accelerated by ram pressure, with radiative cooling playing a crucial role in allowing the gas to survive the acceleration phase \citep{Gronke2018,LiZhihui2019a}.

Dust and molecules provide a powerful discriminant between these ``born comoving'' and ``efficiently pushed'' scenarios. As demonstrated by our results, ram-pressure acceleration of cold gas naturally leads to fast-moving dust and H${}_2$ in the wind. This outcome is difficult to reconcile with the born-comoving scenario, where dust and molecules would be expected to be rapidly destroyed by sputtering in the hot wind before substantial cooling and condensation occur.

In conclusion, the now ubiquitous detection of dusty and molecular outflows \citep[e.g.][]{Walter2017,Leaman2019,Krieger2021} can be naturally interpreted as evidence in favor of efficient acceleration of pre-existing cold gas. In this picture, cold gas is not required to form \emph{in situ} out of the hot wind, but instead survives and is accelerated through a combination of ram pressure and radiative cooling. The presence of dust and molecules at high velocities places particularly strong constraints on the thermal history of the outflow, as both components are highly susceptible to destruction in hot, volume-filling gas. Our results show that efficient acceleration can preserve --- and even regenerate-dust and H${}_2$, providing a self-consistent explanation for the observed multiphase structure of galactic winds.

Our results further show a rapid---and sudden---entrainment of the molecular phase (cf.~Fig.~\ref{fig:fig_H2_sput_vs_growth_time_evol}), occurring on a timescale much shorter than the nominal drag time,
$t_{\rm drag} \sim \chi_{\rm H_2} r / v \sim 300\,(T_{\rm wind}/10^6\,{\rm K})\,t_{\rm cc}$.
We attribute this behavior to two effects: \textit{(i)} efficient momentum transfer to the $\sim 10^4\,{\rm K}$ envelope, which leads to the fiducial entrainment timescale $\sim (T_{\rm wind}/10^4\,{\rm K})^{1/2} t_{\rm cc}$ \citep{Farber2022,Chen2024}, and \textit{(ii)} the preferential formation [destruction] of dust and molecules in [non-]entrained gas.
While the former sets the overall entrainment timescale, the latter is responsible for the sudden ``jump'' in $v({\rm H}_2)$ at $\sim 10$--$15\,t_{\rm cc}$.

This velocity evolution can be directly mapped to observable velocity profiles, which in turn can constrain the mass and momentum transfer between phases. A clear prediction of this work is a---perhaps surprisingly---short molecular entrainment distance and an approximately step-like velocity profile. In practice, variations in clump sizes are expected to smooth out this feature. We note that additional mass transfer from the hot to the colder phase, for example, through externally driven turbulence in galactic winds, can further reduce the entrainment time and distance \citep{Ghosh2026}.

\subsection{The Necessity of Dust Growth for Halo Dust}
Reddening measurements of background sources suggest that galactic halos contain a substantial amount of dust comparable to that in the disks \citep{Menard2010, Peek2015}.
Most of the halo dust resides in the cool ($\sim 10^4$ K) clouds traced by Mg\,{\sc ii} absorbers with DGRs $\sim 0.01$, similar to that of typical galaxies \citep{Menard2012}. 
Our results indicate that dust surviving in the shielded cool clouds is insufficient to account for these high values.
In the entrainment scenario, cool clouds must accrete multiple times their mass from the fast-moving, dust-depleted wind to accelerate to high velocities; hence, the cloud's DGR is inevitably diluted during the process.
Furthermore, in high-temperature winds ($\gtrsim 10^7$~K), the DGR declines orders of magnitude faster than dilution due to sputtering in the cloud-wind interface.
This ``interface sputtering'' continues even after the cloud is fully entrained and presumably remains active throughout its lifetime in the CGM on much longer timescales; e.g., through external turbulence in the CGM (which also causes further cold gas growth and fragmentation; \citealp{Gronke2022}) .
As such, substantial \textit{in situ} dust growth within the entrained clouds is required to reconcile with observations.

Our idealized simulation setup does not account for wind geometry, which may overestimate the effect of sputtering.
In reality, galactic winds expand as they travel away from the disks \citep{Chevalier1985}, leading to altered mass transfer rates even in classic `cloud crushing' setups \citep{Gronke2020,2025MNRAS.544.4621D}.
Galaxy-scale simulations offer complementary insights here despite at lower resolutions.
\citet{Kannan2021} find the DGRs in the hot wind around 60\% and 30\% of the disk values for LMC and Milky Way-mass galaxies, respectively.
In dwarf galaxies,
\citet{Hu2023} find that the wind DGR retains $\sim$80\% of the disk value without dust growth, and can even exceed it when growth is enabled.
\citet{Richie2025} explore different grain sizes and demonstrate that large grains can easily survive in hot winds from starburst galaxies while small grains are rapidly destroyed -- although they neglect the effects of nonthermal sputtering, which can be significant in supernova remnants within the galactic disks \citep{Slavin2015, Hu2019a}. 
The trend of enhanced dust survival in lower mass galaxies can be attributed to lower initial wind densities and temperatures as well as more rapid geometric dilution as the streamlines of the wind diverge and become uncollimated.
Furthermore, as low-mass galaxies typically lack hot gaseous halos, the long-term dust survival is significantly easier than their more massive counterparts.
Therefore, dust carried directly by hot winds likely serves as an efficient source of halo dust in low-mass systems.

Our results also have direct implications for the origin of dust in galaxy clusters.
The observed molecular filaments stretching tens of kpc in the brightest cluster galaxies (BCGs) of cool-core clusters \citep{Salome2006, Russell2017, Russell2019, Olivares2019} strongly indicate the existence of dust.
While condensation directly from the hot intracluster medium (ICM) via thermal instability has been proposed as the source of the cold gas \citep{Sharma2012,McCourt2012,Donahue2022}, it is difficult to explain dust survival in these hot environments.
\citet{Li2019} propose that ejecta from asymptotic giant branch (AGB) stars could seed the cold gas, with dust shielded from the hot gas via efficient radiative cooling.
Our results suggest that as these AGB-seeded clouds interact with the surrounding dust-free hot gas, their DGR would rapidly drop below the value for efficient H$_2$ formation.
Therefore, dust growth is essential to counteract sputtering and dilution in order to explain the observed molecular filaments.

\subsection{PAHs and Molecular Gas in Galactic Winds}

Recent JWST observations have mapped the polycyclic aromatic hydrocarbon (PAH) emission in the galactic wind of M82 in unprecedented detail \citep{Bolatto2024, Fisher2025, Lopez2025}.
In particular, \citet{Lopez2025} shows that PAH emission correlates strongly with cold molecular gas but decouples from the hot, X-ray emitting gas.
In addition, the ratio of PAH emission to neutral gas surface density declines slowly with distance from the disk, mirroring the decline in the radiation field strength.
This suggests that the product of the PAH abundance and the DGR remains surprisingly constant as the wind propagates outward.
Since sputtering reduces both the DGR and the PAH abundance (as small grains are sputtered more rapidly), it is difficult to explain the observed trend in a purely destructive framework.
Dust growth provides a natural mechanism to counteract sputtering, replenishing the dust mass within the entrained clouds and reconciling theory with observations.

To compare the effect of dilution we can relate the mass doubling time of the cloud
\begin{align}
    t_{2m} \equiv& \frac{m}{\dot m}\sim \frac{\chi V_c}{A_c c_{\rm s,c}}\sim \chi \mathcal{M} \left(\frac{r}{L}\right)^{1/2} t_{\rm cc}
\end{align}
where in the last step we made the simplifying assumption\footnote{Note that we also set $v_{\rm mix}\sim c_{\rm s,c}$ and omitted the factor $(t_{\rm cool}/t_{\rm sc})^{1/4}$ \citep{Gronke2020}.} that the cloud is stretching to a cylinder with (tail) length $L$ which we can estimate to be $L\sim \chi r$ yielding $t_{2m}\sim \chi^{1/2}t_{\rm sc,cl}$ (in agreement with Fig.~\ref{fig:fig_warmcold_time_evol_single} albeit our high $\chi$ runs seem to have an even larger area-to-volume ratio), i.e., $\sim $a few Myr for our fiducial values.

Observations of molecular outflows have commonly been interpreted as a suppressing mechanism for star formation by directly removing the molecular reservoir within the disks \citep{Feruglio2010, Sturm2011, Bolatto2013a, Cicone2014, Spilker2020}.
However, our results suggest that the high-velocity molecular gas may form \textit{in-situ} in the entrained clouds that were initially warm ($\sim 10^4$~K) and atomic. 
Consequently, the observed molecular outflows may be entirely distinct from the star-forming molecular clouds in galactic disks.

\subsection{Potential Caveats}

Our dust evolution model only includes processes that exchange mass between dust and gas, i.e., sputtering and dust growth.
In addition, we adopt a fixed grain size distribution, 
which implicitly assumes that other dust processes (e.g., shattering) are keeping the distribution constant.
Finally, our dust growth timescale (Eq.~\ref{eq:t_grow}) is based on the physical collision rate, which can be affected by several factors we neglect, such as Coulomb focusing and the uncertainty of the sticking coefficient \citep[e.g.][]{Weingartner1999, Zhukovska2014}.
We plan to revisit the problem with a more sophisticated dust evolution model.

The DGR and H$_2$ fraction in our simulations are still not numerically converged (cf. Appendix \ref{app:conv}).
This is primarily due to the difficulty of resolving the spatially compact cold phase of the cloud.
Therefore, properties associated with the cold phase are more difficult to converge than the global properties of the cloud.
However, based on the trend in the convergence tests, we expect the effect of dust growth to be even more significant in a fully resolved simulation.
Apart from dust physics, we also ignored other microphysical effects which in principle can change the evolution of multiphase gas -- such as thermal conduction and viscosity. However, previous work has shown that in such systems, the mass growth is dominated by turbulent diffusion and is not altered significantly by such `laminar' effects \citep{Tan2021,Bruggen2023,Marin-Gilabert2026}. We thus do not expect conduction and viscosity to change our results significantly.

Finally, our idealized setup neglects global geometry.
In reality, galactic winds expand spherically as they travel away from the disk, which can suppress cloud growth and the development of the two-phase structure \citep{2025MNRAS.544.4621D}, thereby affecting the formation of molecular gas. 
Similarly, the initial state of a spherical cold gas cloud does not resemble a more complex ISM. 
Recently, \citet{Hidalgo-Pineda2026} (also see, e.g., \citealp{Banda-Barragan2020a}) studied the interaction of a hot wind with a more complex ISM morphology and found that while cold gas can also survive under certain conditions, there is additional turbulence seeded in the wind leading to cold gas fragmentation (in agreement with larger scale simulations, see, e.g., \citealp{Schneider2020,Steinwandel2024}). This naturally affects shielding of the cold gas and thus dust and molecular growth.
Generally, high-resolution global simulations resolving the cold molecular phase are needed to obtain a more complete picture.

\section{Summary}
\label{sec:conclusion}

We have performed the first cloud-crushing simulations coupled with dust evolution (including sputtering and growth) \textit{and} a time-dependent chemistry network.
This enables us to systematically investigate the conditions for dust survival and H$_2$ formation in galactic outflows.
To this end, we have run various combinations of initial cloud densities ($n^\prime_{\rm c} = $ 3, 10, and 30), wind temperatures ($T_{\rm w} = 10^6$ K and $10^7$ K), and dust evolution processes.
Our main findings can be summarized as follows.

\begin{enumerate}

	\item     
    Clouds that survive the cloud-wind interaction ($n^\prime_{\rm c} > 1$) naturally develop into a two-phase structure characterized by a cold ($\sim 30$ K), dense core embedded in a warm ($\sim 10^4$ K), diffuse envelope as long as low-temperature cooling is enabled.
    The cold phase is about 20 -- 200 times denser than the initial cloud density.
    However, the cold phase can still be HI-dominated if its density or DGR is too low.

    \item 
    In $10^6$~K winds, dust initially in the cloud (the ``cloud dust'') remains largely intact and well-shielded from the hot gas as long as the cloud is in the cooling-dominated regime ($n^\prime_{\rm c} > 1$).
    However, the survival of the cloud dust does not guarantee molecular winds; the accretion of the dust-depleted wind material significantly dilutes the cloud's DGR to a level insufficient for H$_2$ formation.
    The key to producing molecular winds lies in dust growth, which can outcompete this dilution and eventually recover the original DGR.
    This, however, requires cloud densities significantly higher than the threshold density ($n^\prime_{\rm c} \gtrsim 30$) to facilitate efficient dust growth.
    Lower-density clouds remain dust-poor and atomic at all times.

    \item 
    In $10^7$\,K winds, the cloud's DGR decreases much more rapidly than dilution alone can account for.
    This implies that cloud dust is substantially sputtered within the cloud-wind interface where temperatures are now sufficiently high to trigger sputtering. 
    At $n^\prime_{\rm c} = 3$, this leads to essentially dust-free clouds dominated by atomic gas.
    In high-density clouds ($n^\prime_{\rm c} \gtrsim 10$), dust growth can ``undo'' the damage by sputtering and recover the original DGR, providing the necessary conditions for molecular winds.

    \item 
    Nonthermal sputtering plays a subdominant role in dust destruction, as there is no mechanism to continuously decouple dust and gas after entrainment.
    In contrast, thermal sputtering remains active as long as the interface temperatures are sufficiently high.

    \item 
    High-velocity molecular gas can form \textit{in-situ} in entrained clouds that were initially warm ($\sim 10^4$~K), atomic, and not star-forming. 
    Consequently, the observed molecular outflows may be entirely distinct from the star-forming molecular clouds in galactic disks.

\end{enumerate}

Our results suggest that \textit{in situ} dust growth is essential to explain the observed abundance of dust in the CGM and the dusty, molecular gas in galactic winds.
Future studies that include more comprehensive physical processes and more realistic initial conditions are needed.

\section*{Acknowledgments}
C.Y.H. acknowledges support from the National Science and Technology Council (NSTC) of Taiwan under grant 113-2112-M-002-041-MY3.
C.Y.H. thanks the computational and storage resources provided by the National Center for High-Performance Computing (NCHC) of the National Institutes of Applied Research (NIAR) 
as well as 
the Academia Sinica Grid-computing Center (ASGC) supported by Academia Sinica.
MG thanks the European Union for support through ERC-2024-STG 101165038 (ReMMU).

\bibliography{literatur,refs_extra}{}
\bibliographystyle{aasjournal}

\appendix

\begin{figure*}
	\centering
	\includegraphics[width=0.9\linewidth]{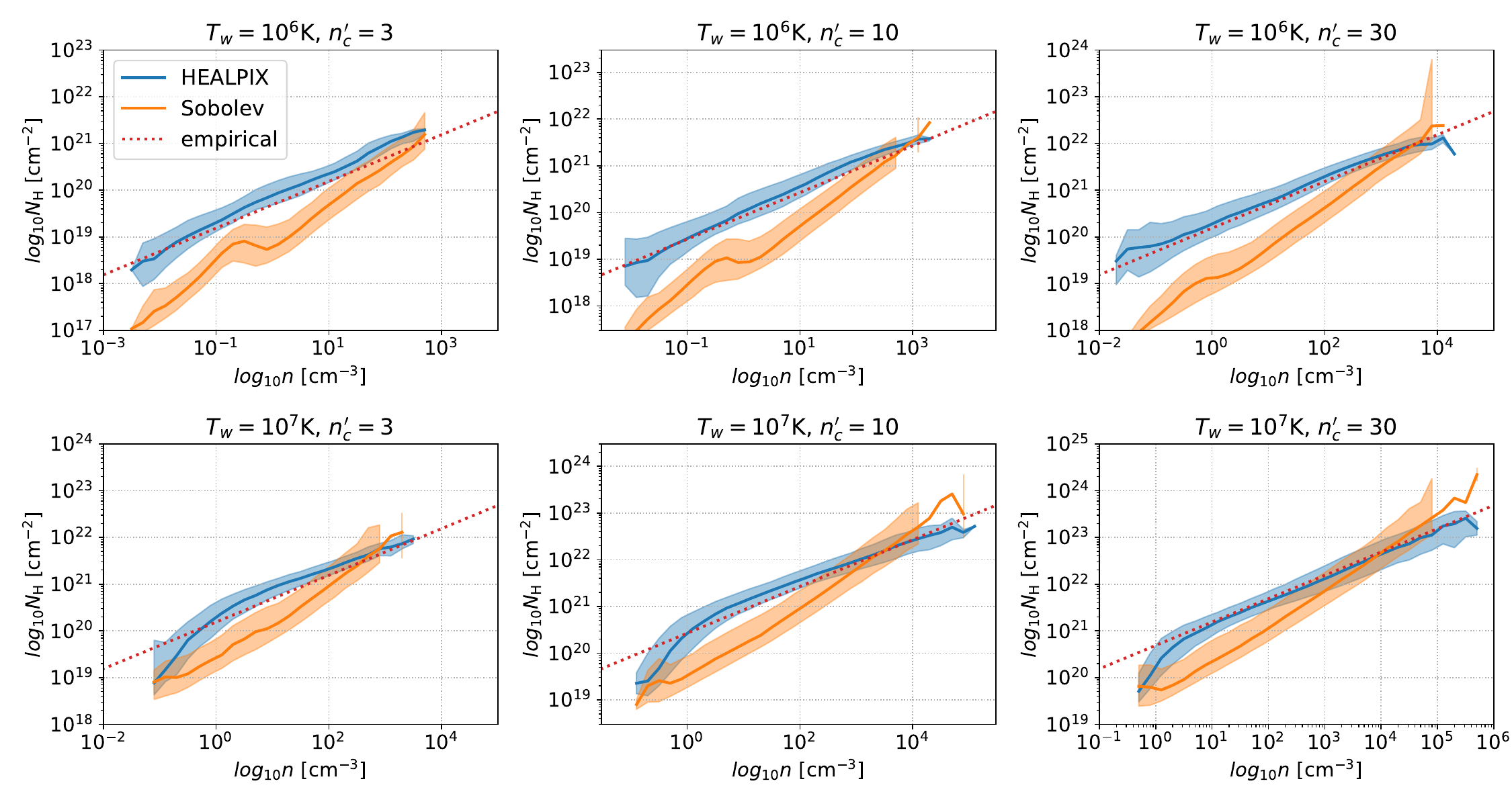}
	\caption{
    The relationship between the volumetric density $n$ and the column density for shielding against the FUV radiation field $N_{\rm H}$.
    The blue solid lines show the median column density at a given density obtained via the {\sc HEALPix}-based method and the shaded areas represent the 25 and 75 percentiles.
    Similarly, the results obtained by the ``Sobolev approximation'' are shown in orange.
    The HEALPix-based method closely follows an empirical relation (Eq.~\ref{eq:NH}) shown by the red dotted lines.
    In contrast, the Sobolev approximation systematically underestimates column densities.
    }
	\label{fig:nhvsnh}
\end{figure*}

\section{Empirical Relation between Density and Column Density}\label{app:NHnH}

In hydrodynamical simulations coupled with chemistry, 
a common method to obtain the column density for shielding against the FUV radiation field is the {\sc TreeCol} algorithm \citep{Clark2012},
where the column density is calculated by averaging over 12 different sightlines defined by the {\sc HEALPix} algorithm \citep{Gorski2011},
typically tied with the gravity solver to leverage the tree structure and minimize the computational overhead \citep{Clark2012, Walch2015a, Hu2016, Hu2021a}.
However, 
as our code currently does not support this functionality for an elongated simulation box, 
we adopt an empirical relationship between the volumetric density $n$ and the column density:
\begin{eqnarray}\label{eq:NH}
N_{\rm H} = 0.5 n_c r_{\rm c} \left( \frac{\chi}{100}  \right)^{-0.5} \left( \frac{n}{n_c}  \right)^{0.5},
\end{eqnarray} 
established by post-processing our simulations using the same algorithm.
We integrate the column densities up to a fixed radius of $r_{\rm c}$, the characteristic length of the cloud.
The results are shown in Fig.~\ref{fig:nhvsnh}. 
Our empirical relation (red dotted lines) closely follows the {\sc HEALPix}-based column densities (in blue). 
In contrast,
the widely adopted ``Sobolev approximation'' (in orange),
$N_{\rm H} = n / \nabla n$,
systematically underestimates column densities in all cases.

\section{Convergence Study}\label{app:conv}
\begin{figure*}
    \centering
    \includegraphics[width=1\linewidth]{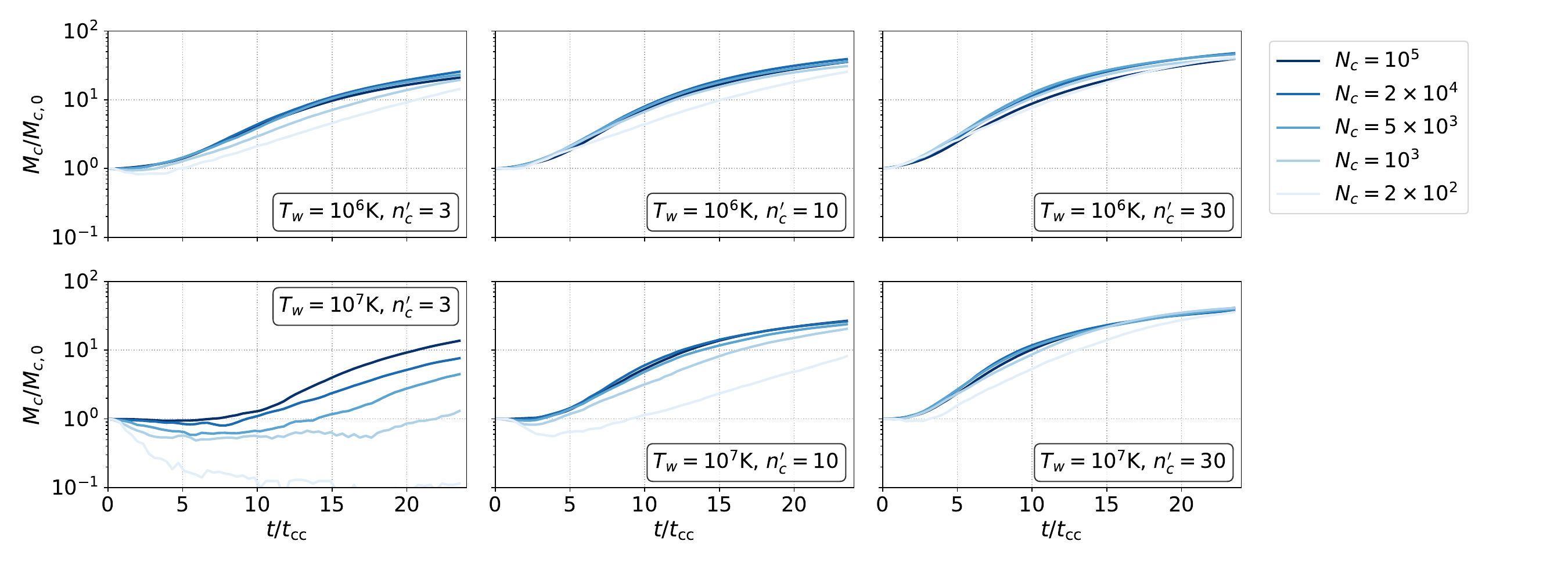}
    \caption{
    Convergence test for the normalized cloud mass (the current cloud mass divided by the initial cloud mass) as a function of time.
    }
    \label{fig:conv_Mc_time_evol_single}
\end{figure*}

\begin{figure*}
    \centering
    \includegraphics[width=1\linewidth]{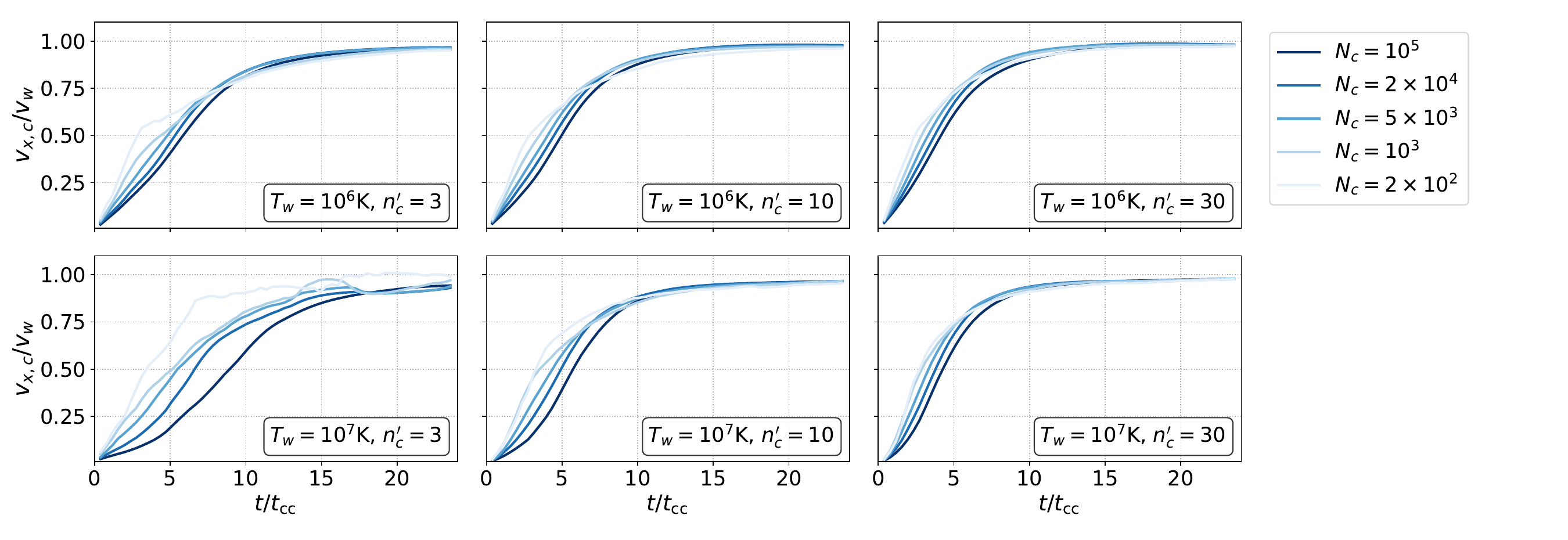}
    \caption{
    Convergence test for the normalized cloud velocity (mass-weighted cloud velocity along the wind direction divided by the wind velocity)  as a function of time.
    }
    \label{fig:conv_vxc_time_evol_single}
\end{figure*}

\begin{figure*}
    \centering
    \includegraphics[width=1\linewidth]{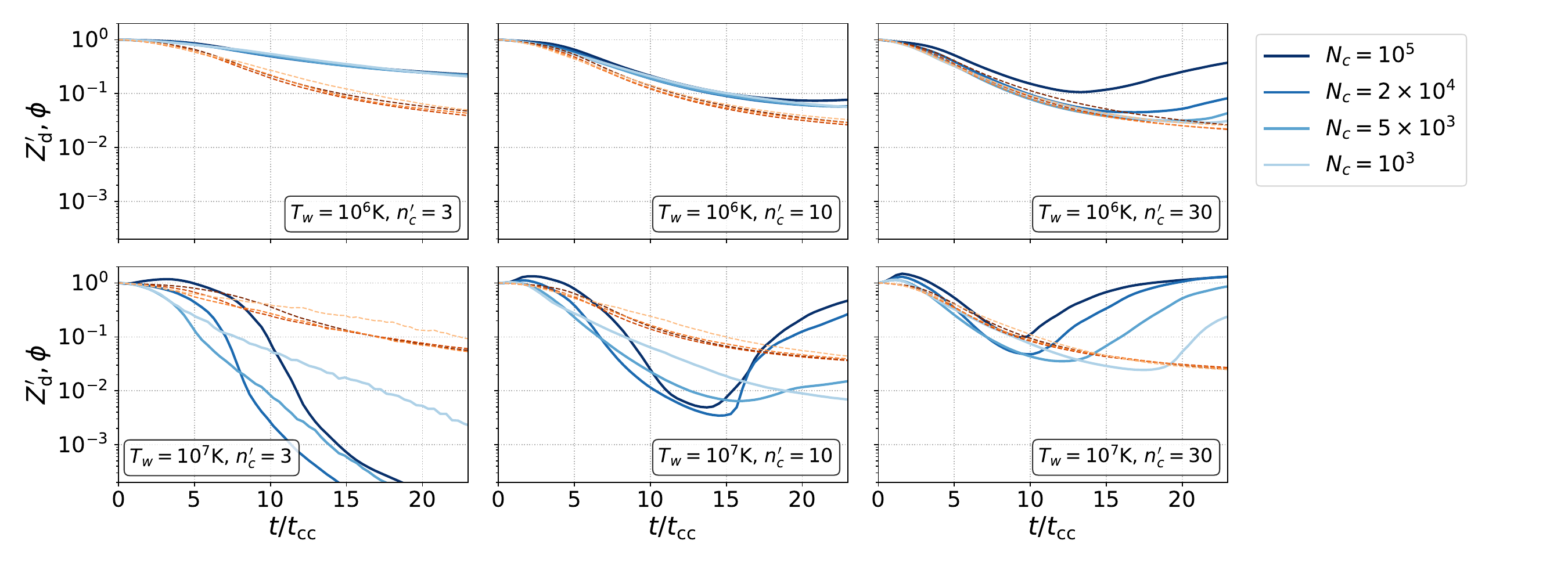}
    \caption{
    Convergence test for the DGR of the cloud normalized to the Milky Way value (blue solid lines) as a function of time. 
    The abundance of the passive scalar tracing the initial cloud material is shown in orange dotted lines. 
    }
    \label{fig:conv_dgr_sputter_vs_growth_time_evol_all}
\end{figure*}

\begin{figure*}
    \centering
    \includegraphics[width=1\linewidth]{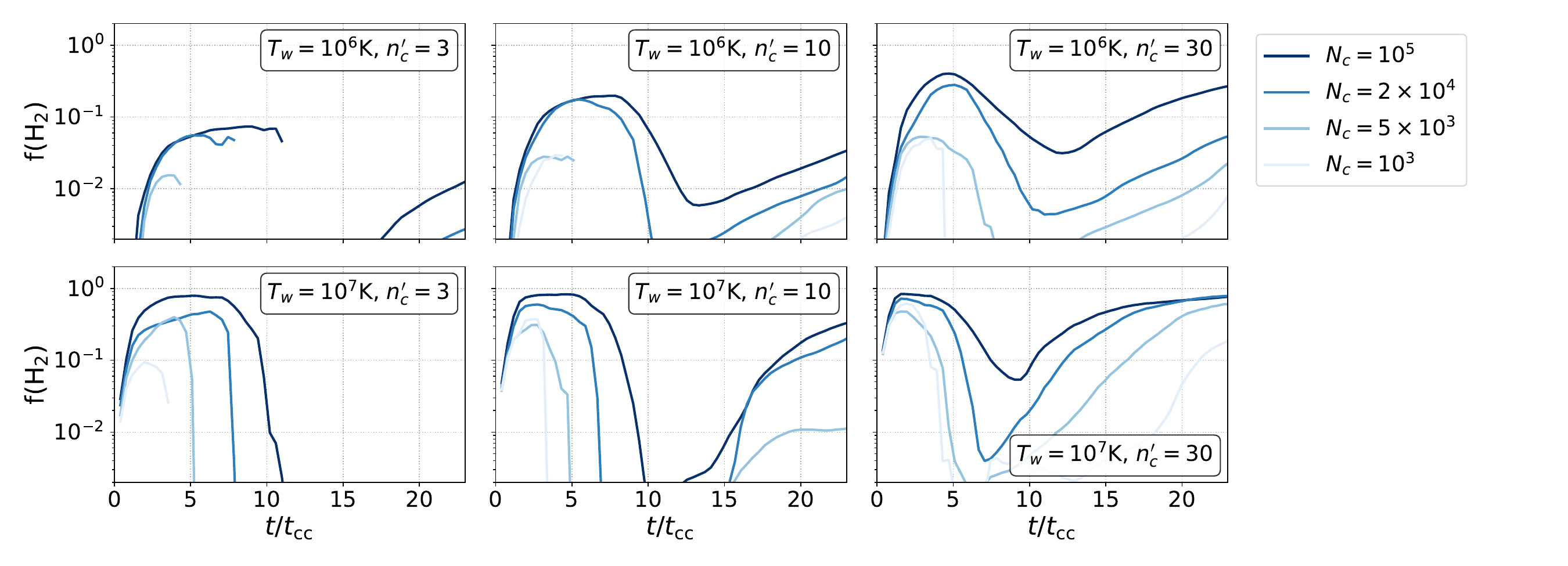}
    \caption{
    Convergence test for the H$_2$ mass fraction of the cloud as a function of time.
    }
    \label{fig:conv_H2_all_time_evol}
\end{figure*}

In this section,
we present convergence tests 
for the time evolution of the following cloud properties:
\begin{enumerate}
    \item Normalized cloud mass (Fig.~\ref{fig:conv_Mc_time_evol_single}).
    \item Cloud velocity normalized to the wind velocity (Fig.~\ref{fig:conv_vxc_time_evol_single}).
    \item DGR of the cloud normalized to the Milky Way value (Fig.~\ref{fig:conv_dgr_sputter_vs_growth_time_evol_all}).
    \item H$_2$ mass fraction of the cloud (Fig.~\ref{fig:conv_H2_all_time_evol}).
\end{enumerate}
The number of particles in the cloud at $t = 0$ is systematically varied from $N = 10^5, 2\times 10^4, 5\times 10^3, 10^3$, and $2\times 10^2$,
which correspond to resolving the cloud radius with 29, 17, 11, 6.3, and 3.7 cells, respectively.

The cloud mass and cloud velocity are both well converged.
In almost all cases,
convergence is achieved at $N_{\rm c} = 5\times 10^3$.
Even the runs with $N_{\rm c} = 200$ show reasonable agreement with the converged runs.
This is consistent with \citet{Gronke2020} who showed that the resolution requirement is surprisingly modest ---
8 cells are sufficient to capture the mass growth for Eulerian codes.
The $T_{\rm w} = 10^7$~K and $n^\prime_{\rm c} = 3$ run shows a stronger resolution dependence,
presumably because the system is in transition from mixing-dominated to cooling-dominated and is thus more sensitive to numerics.

In contrast, the convergence on the cloud's DGR is more complicated.
For the $T_{\rm c} = 10^6$~K wind, the DGR converges well in both $n^\prime_{\rm c} = 3$ and $n^\prime_{\rm c} = 10$ cases. 
Since the DGR in these runs is governed by the dilution effect, which is naturally converged as the mass growth is converged.
However, the DGR shows a strong resolution dependence in the $n^\prime_{\rm c} = 30$ case and remains unconverged even at $N_{\rm c} = 10^5$, and the low-resolution runs significantly underestimate dust growth.
On the other hand, the passive scalar is already converged.
This indicates that while the cloud mass is converged, the internal two-phase structure of the cloud is not.
Since the cold phase of the cloud (where dust growth occurs) is typically dense and compact, it is significantly more difficult to resolve than the mass growth.

For the $T_{\rm c} = 10^7$~K wind, while the passive scalar is converged, the cloud's DGR is unconverged in all three cases.
For $n^\prime_{\rm c} = 10$ and $n^\prime_{\rm c} = 30$ cases, this is primarily due to the difficulty to resolve the cold phase, which in turn underestimates dust growth at low resolutions.
However, the unconverged DGR in the $n^\prime_{\rm c} = 3$ case cannot be explained by not resolving the cold phase, as dust growth is inefficient here.
Instead, it is likely because the temperature distribution in the cloud-wind interface is more difficult to converge \citep{Tan2021}.
Since thermal sputtering is only efficient at $T\gtrsim10^6$~K, dust destruction is sensitive to the interface temperature distribution and thus remains unconverged.

Similarly, the H$_2$ fraction of the cloud is highly sensitive to resolution, 
even in cases where dust growth is unimportant and the DGR is converged.
As the low-resolution runs fail to resolve the cold phase,
not only is dust growth strongly suppressed,
but also the cold-gas densities are underestimated,
both of which act against H$_2$ formation.

In summary,
while cloud mass and cloud velocity are well converged,
the DGR and H$_2$ fraction are not and should be viewed as lower limits.
This further supports our conclusion on the importance of dust growth, which should play a more significant role when the internal structure of the cloud is fully resolved.

\end{document}